\documentclass[prd,showpacs,preprint,eqsecnum,amsmath,amssymb,nofootinbib]{revtex4}
\usepackage{graphicx, tikz}
\usepackage{amssymb,amsmath, mathrsfs}
\usepackage{amssymb, graphics, setspace}
\usepackage{hyperref}
\usepackage[all]{xy}
\newcommand{\be}{\begin{equation}}
\newcommand{\ee}{\end{equation}}
\newcommand{\bea}{\begin{eqnarray}}
\newcommand{\eea}{\end{eqnarray}}
\newcommand{\bes}{\begin{subequations}}
\newcommand{\ees}{\end{subequations}}

\newcommand{\w}{\omega}
\begin{document}
\title{Particle Production in the Interiors of Acoustic Black Holes}
\author{Roberto~Balbinot}
\email{balbinot@bo.infn.it}
\affiliation{Dipartimento di Fisica dell'Universit\`a di Bologna and INFN sezione di Bologna, Via Irnerio 46, 40126 Bologna, Italy\\
Centro Fermi - Museo Storico della Fisica e Centro Studi e Ricerche Enrico Fermi, Piazza del Viminale 1, 00184 Roma, Italy}
\author{Alessandro~Fabbri}
\email{afabbri@ific.uv.es}
\affiliation{Departamento de F\'isica Te\'orica and IFIC, Universidad de Valencia-CSIC, C. Dr. Moliner 50, 46100 Burjassot, Spain\\  Centro Fermi - Museo Storico della Fisica e Centro Studi e Ricerche Enrico Fermi, Piazza del Viminale 1, 00184 Roma, Italy  \\
Laboratoire de Physique Th\'eorique, CNRS UMR 8627, B\^at. 210, Universit\'e Paris-Sud 11, 91405 Orsay Cedex, France}
\author{Richard~A.~Dudley}
\email{dudlra13@wfu.edu}
\author{Paul~R.~Anderson}
\email{anderson@wfu.edu}
\affiliation{Department of Physics, Wake Forest University, Winston-Salem, North Carolina 27109, USA}
\begin{abstract}
Phonon creation inside the horizons of acoustic black holes is investigated using two simple toy models.
It is shown that, unlike what occurs in the exterior regions, the spectrum is not thermal. This non-thermality is due to the anomalous scattering that occurs in the interior regions.
\end{abstract}
\maketitle
\section{Introduction}

Among the many spectacular celestial objects that populate our universe, black holes (BHs) are perhaps the most intriguing.  The presence of a causal horizon prevents any direct observation of the interior.  According to theoretical studies based on General Relativity, the interior is a place where peculiar physical effects occur, which cannot be confirmed by astronomical observations.  The situation has positively changed in recent times with the advent of so called ``analog BHs'' \cite{Unruh,Barcelo:2005fc}.  These are condensed matter systems that are realizable in the laboratory, which mimic some of the essential features of gravitational BHs. A typical example is a Bose-Einstein Condensate (BEC) fluid (see for example  \cite{ps}) whose flow becomes supersonic \cite{garayetal, Macher:2009nz, rpc}. The supersonic region, trapping sound waves inside it, is the analog of the BH interior.  The sonic surface where the speed of the flow equals the local speed of sound, plays the role of the horizon. This sonic horizon however has no causal significance at all: there is nothing  to prevent one from directly observing the interior region.  Indeed the first experimental observations of the analog of Hawking radiation~\cite{hawking} in BECs by Steinhauer et.\ al.~\cite{jeff1, jeff2} were made by performing simultaneous measurements of the density outside and inside the sonic horizon.  A peak was observed in the resulting in-out density-density correlation function that was predicted in~\cite{paper1, paper2} and which is the `smoking gun' signaling the presence of Hawking radiation.  In the same spirit, one can imagine that other processes that are predicted to take place in the interior of a BH can be experimentally verified by looking at appropriate analog models.

With this as motivation, in this paper we discuss the unusual features of scattering by a potential inside the horizon of a stationary BEC analog BH and its consequences.
The calculations are done in the analog spacetime using quantum field theory in curved space techniques. These are the same types of calculations that one would do to explore similar effects in the interior of a real black hole.

\begin{figure}
\begin{tikzpicture}
\draw[->]   (4.5,1) -- node[above] {Incident} (2.25,1);
\draw[->]   (-2.25,1.5) --node[above] {T} (-4.5,1.5);
\draw[->]   (2.25,2)--node[above] {R} (4.5,2) ;
\draw[-]   (-4,0) -- (4,0);
\draw [cyan] plot [smooth] coordinates { (-4,0) (-2,0.5) (0,3)  (2,0.5) (4,0)};
\end{tikzpicture}
\caption{\label{Fig:PlaneWaveIncident1}Illustration of a  plane wave incident onto a potential from the right, which then is partially  transmitted  to the left and also partial reflected back to the right of the barrier.
}
\end{figure}
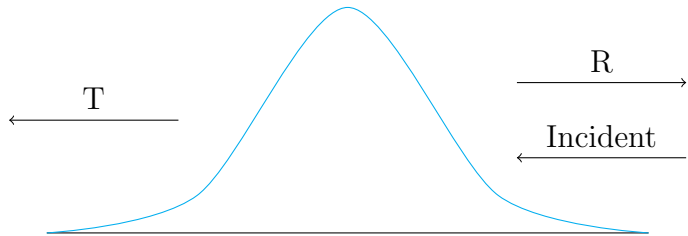

In Quantum Mechanics in the presence of a potential, an  incident flux is split into a transmitted and a reflected part (see Fig.\ref{Fig:PlaneWaveIncident1}).
Reflection ($R$) and Transmission ($T$) coefficients satisfy the unitary relation $|R|^2+|T|^2=1$, which is the conservation of probability. Note that the previous relation implies that $|R|^2 $ and $|T|^2 \leq 1$.

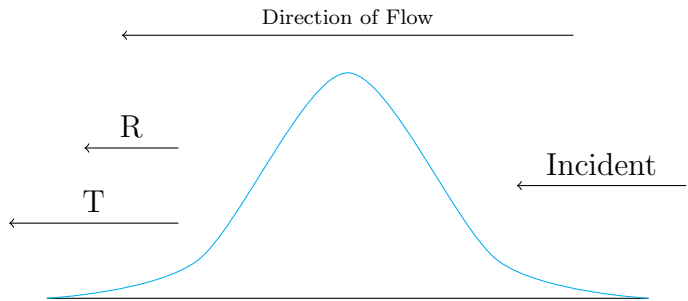
\begin{figure}
\begin{tikzpicture}
\draw[->]   (4.5,1.5) -- node[above] {Incident} (2.25,1.5);
\draw[->]   (-2.25,1) --node[above] {T} (-4.5,1);
\draw[->]   (-2.25,2)--node[above] {R} (-3.5,2) ;
\draw[-]   (-4,0) -- (4,0);
\draw [cyan] plot [smooth] coordinates { (-4,0) (-2,0.5) (0,3)  (2,0.5) (4,0)};
\draw[->]   (3,3.5) -- node[above] {\scriptsize{Direction of Flow}} (-3,3.5);
\end{tikzpicture}
\caption{\label{Fig:PlaneWaveIncident2}Illustration of a  plane wave incident onto a potential from the right in the interior of a BH, which then is partially  transmitted  to the left and also partial reflected, however the reflected portion is also moving to the left of the barrier since in the interior the wave is forced to travel further into the BH.}
\end{figure}

Inside the horizon of a BH both  the transmitted and the ``would be reflected'' part of the field are forced to propagate in the same direction, namely towards the center of  the black hole (see Fig. \ref{Fig:PlaneWaveIncident2}).
The scattering is `anomalous' and $R$ and $T$ no longer satisfy the previous unitary relation. Instead they satisfy $|T|^2-|R|^2=1$ which implies particle creation since $|T|^2\geq 1$.  Another way to think about this is that, while the outside region of a nonrotating black hole is static, the interior can be thought of as a dynamical cosmology in which particle creation occurs.

 We shall deal with both massless and massive quantum fields. For the latter case there is usually a mass gap, namely $E\geq m$ where $E$ is the conserved (Killing) energy and $m$ is the mass of the particle. Inside a BH the former inequality no longer holds, $E$ can take any value, even negative ones.

In Sec. II a brief review is given of the set-up for BEC analog black holes.  In Sec. III particle production is investigated in the case of massless phonons with a double delta function potential.  In Sec. IV particle production is investigated for massive phonons when the effective potential is zero and the mass term in the mode equation is approximated by two step functions.  Sec. V contains a discussion of our results and comparisons with some previous work.

\section{The Setting}
Under the hydrodynamic approximation the phase fluctuation operator $\hat{\phi}$ in a BEC satisfies a covariant version of D`Alembert's wave equation (see for instance \cite{Barcelo:2005fc})
\be
\hat{\Box}\ \hat{\phi}=0\ ,\label{Eq:WaveEquation1}
\ee
where $\hat{\Box}=\hat{\nabla}_\mu\hat{\nabla}^\nu$ is evaluated on a fictitious curved spacetime metric, called the acoustic metric, which in our case we write as follows
\be
ds^2=\frac{n}{m_ac}\left[-c^2(x)dT^2+\left(dx+v_0dT\right)^2+dy^2+dz^2\right]\label{Eq:LineElement1}
\ee
where $n$ is the density of the condensate (here assumed to be constant), $m_a$ is the mass of a single atom of the BEC, and $c(x)$ is the sound speed.  The flow is assumed to be stationary and one dimensional along the $x$ axis with the velocity $\vec{v}=-v_o\hat{x}$ constant and directed from right to left.

For a typical profile used in BEC analog models $c(x)$ becomes constant in both asymptotic regions ($x\to\pm \infty$) so that $\lim_{x\to +\infty}c(x)=c_r$ and $\lim_{x\to -\infty}c(x)=c_l$ with $c_r>v_0$ and $c_l<v_0$.  Thus  the asymptotic regions are homogeneous and the profile monotonically decreases from right to left. The profile $c(x)$ is chosen so that the horizon $c(x)=v_0$ is at $x=0$.  In the region $x<0$, where $c(x)<v_0$ the metric describes the interior region of the acoustic BH while for $x >0$, where $c(x)>v_0$, the metric describes the exterior region of the acoustic BH. We call the exterior the $r$ region and the interior the $l$ region.
Performing a dimensional reduction along the transverse direction and passing from the Gullstrand-Painlev\'{e} coordinates ($T, x$) to the Schwarzschild like ones ($t, x^*$) via the transformation
\be
t=T-\int dx \frac{v_0}{c(x)^2-v_0^2} \quad \text{and}\quad  x^*=\int  dx \frac{c(x)}{c(x)^2-v_0^2} \ . \label{xstarandt}
\ee
the wave equation (\ref{Eq:WaveEquation1}) can be reduced to
\be
 [ -\partial_t^2 +\partial_{x^*}^2 -k_\perp^2(c^2-v_0^2)+V_{\text{eff}} ]\hat{\phi}^{(2)} =0\ ,  \label{Eq:WaveEquation2}
\ee
where the effective potential is given by
\be
V_\text{eff}\equiv\frac{c^2-v_0^2}{c}\left[ \frac{1}{2}\frac{d^2c}{dx^2}\left(1-\frac{v_0^2}{c^2}\right)-\frac{1}{4c}\left(\frac{dc}{dx}\right)^2+\frac{5v_0^2}{4c^3}\left(\frac{dc}{dx}\right)^2\right] \ .
\ee
The coefficient $k^2_\perp$ is related to the transverse momentum and $\hat{\phi}^{(2)}$ is the dimensionally reduced field operator (see the appendix of Ref. \cite{massive2} for details).
The last two terms in Eq. (\ref{Eq:WaveEquation2}), the mass-like term and $V_{\text{eff}}$, cause scattering of the modes. Note that both of these terms vanish at the horizon. There the modes are effectively massless and propagate freely. The second coordinate transformation in Eq. (\ref{xstarandt}) maps the $(0, +\infty )$ interval in x in the $r$ region to $(-\infty, +\infty)$ in $x^*$ while in $l$ the interval $(-\infty,0 )$ in $x$ is mapped to $(+\infty, -\infty)$ in $x^*$.

According to the standard procedure of quantum field theory in curved space-time, the field operator $\hat{\phi}^{(2)}$ is expanded in terms of a complete set of basis functions  $\{f_\w, f_\w^*\}$, which are solutions of the classical counterpart of the operator equation (\ref{Eq:WaveEquation2}) with the result
\be
\hat{\phi}^{(2)}= \int_0^\infty d \w \; \left(\hat{a}_\w f_\w + \hat{a}_\w^\dagger f_\w^{*} \right)\label{Eq:ModeExpansion1}
\ee
The creation and annihilation operators, $\hat{a}_\w$ and $\hat{a}_\w^\dagger$, satisfy the usual commutation relations. The modes $f_\w$ are normalized using the conserved scalar product
\be
 \label{sc} (f_\w,f_{\w'})= -i\int d\Sigma^\mu f_\w\overleftrightarrow{\partial_\mu} f_{\w'}^{*} [g_\Sigma (x)]^{\frac{1}{2}}
   \ee
with $d\Sigma^\mu=n^\mu d\Sigma$, where $\Sigma$ is a Cauchy surface, $n^\mu$ a future directed unit vector perpendicular to $\Sigma$, and $g_\Sigma$ the determinant of the induced metric.
Writing
\be \label{uu}  f_\w = e^{\pm i \w t} \chi_\w(x^*)   \ee
and substituting into~\eqref{Eq:WaveEquation2} gives
\be \frac{d^2 \chi_\w}{d x^{* \;2}} + \left( \w^2 - k_\perp^2 (c^2 - v_0^2) + V_{\rm eff}\right) \chi_\w = 0 \;. \label{Eq:WaveEquation3} \ee

In this paper we consider two toy models for the terms in Eq. (\ref{Eq:WaveEquation3}) responsible for the scattering which have the advantage of being exactly solvable  while, despite their crudeness, encode all of the basic features of the process we wish to discuss.

\section{Dirac Delta Function Potentials\label{Sec:DiracDeltaFunctionPotential}}

\begin{figure}
\begin{tikzpicture}
\node (I)    at ( 4,0)   {$r$};
\node (II)   at (-4,0)   {};
\node (III)  at (0, 2.5) {};
\node (IV)   at (0,-2.5) {};
\node (V)   at (2.,-2.) {};
\node (VI)   at (6.25,-1.75) {};
\node (VII)    at ( 4.,2.5)   {};
\node (VIII)   at (-2.,2.) {};

\node (IX)    at ( 0,4)   {$l$};

\path 
   (IX) +(90:4)  coordinate  (IXtop)
       +(-90:4) coordinate  (IXbot)
       +(180:4) coordinate (IXleft)
       +(0:4)   coordinate (IXright)
       ;
\draw (IXleft) --
          node[midway, below, sloped] {$_v I_+^l$}
      (IXtop) --
          node[midway, above right]    {$_u I_+^l$}
          node[midway, below, sloped] {}
      (IXright) --
          node[midway, below right]    {}
          node[midway, above, sloped] {}
      (IXbot) --
          node[midway, above, sloped] {}
      (IXleft) -- cycle;

\path  
  (II) +(90:4)  coordinate[label=90:]  (IItop)
       +(0:4)   coordinate                  (IIright)
       ;
\draw
      (IItop) --
          node[midway, below, sloped] {$H_-$}
      (IIright)-- cycle;

\path 
   (I) +(90:4)  coordinate[label=90:$\quad i_+$]  (Itop)
       +(-90:4) coordinate[label=-90:$i_-$]  (Ibot)
       +(180:4) coordinate (Ileft)
       +(0:4)   coordinate[label=0:$i_0$] (Iright)
       ;
\draw (Ileft) --
          node[midway, below, sloped] {$\quad \quad H_+$}
      (Itop) --
          node[midway, above right]    {$ I_+^r$}
          node[midway, below, sloped] {}
      (Iright) --
          node[midway, below right]    {$ I_-^r$}
          node[midway, above, sloped] {}
      (Ibot) --
          node[midway, above, sloped] {}
      (Ileft) -- cycle;
\draw  (Ileft) -- (Itop) -- (Iright) -- (Ibot) -- (Ileft) -- cycle;

{\color{blue}\path[->]
 (V) +(45:0) coordinate[label=45:$\ \quad ^{in}\!f_H^r$](IVtop)
	+(45:2) coordinate(IVbot)

	;
\draw[->] (IVtop) ->(IVbot);

\path[->]
 (VI) +(45:0) coordinate[](IVOut)
	+(135:1.5) coordinate(IVIn)

;
\draw[->] (IVOut) ->(IVIn);

\path[->]
 (VI) +(90:.2) coordinate[label=90:$ ^{in}\!f_I^r  $ ](IVOut2)
	+(135:1.5) coordinate(IVIn2);
	
\path[->]
 (VIII) +(45:0) coordinate[label=45:$\ \quad ^{in}\!f_H^l$](VIIItop)
	+(45:1.5) coordinate(VIIIbot)
;
\draw[->] (VIIItop) ->(VIIIbot);
}
\end{tikzpicture}
\caption{\label{Fig:Penrose1} Penrose diagram with the ``in'' mode basis schematically illustrated  in the ${l}$ and ${r}$ regions.}
\end{figure}
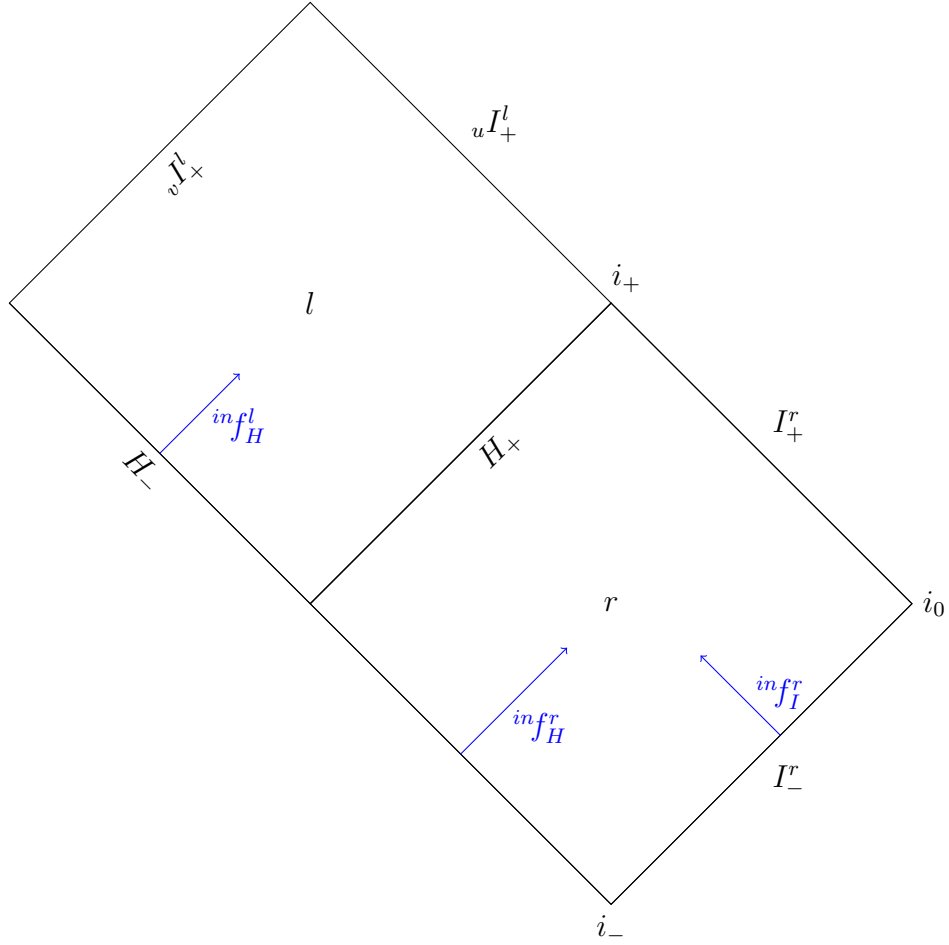

In the first toy model, the transverse excitations are neglected (i.e., $k_\perp=0$) and $V_\text{eff}$ is approximated by two Dirac delta functions, one in region $r$ and one in region $l$. For simplicity we choose them at $x_r^*=0$ in $r$ and at $x_l^*=0$ in $l$ leading to \footnote{For typical flows discussed in the literature which mimic the experimental set up in Ref. \cite{jeff1, jeff2} the effective potential in the interior is dominated by a negative peak. Thus, while our analytic results are valid for
arbitrary values of $V_l$, when plotting the results we restrict our attention to the case $V_l < 0$.}
\begin{align}
V_{\text{eff}}= \left\{
        \begin{array}{ll}
            V_l\delta (x^*-x^*_l)\ , & \quad x < 0\ , \\
           V_r\delta (x^*-x^*_r)\ , & \quad x > 0\ .
        \end{array}
    \right.   \label{2-delta}
\end{align}
The Penrose diagram for the BH metric given in Eq. (\ref{Eq:LineElement1}) is shown in Fig \ref{Fig:Penrose1}, where the modes  representing our `in' basis are schematically indicated.

The asymptotic behaviors of these modes are
\be
^{in}\!f_I^r=\frac{e^{-i\omega t}e^{-i\omega x^*}}{\sqrt{4\pi \omega}}=\frac{e^{-i\omega v}}{\sqrt{4\pi \omega}}\label{Eq:ModeDef1}
\ee
on past null infinity, $I^r_-$;
\be
^{in}\!f_H^r=\frac{e^{-i\omega t}e^{i\omega x^*}}{\sqrt{4\pi \omega}}=\frac{e^{-i\omega u}}{\sqrt{4\pi \omega}}\label{Eq:ModeDef2}
\ee
on the portion of the past horizon in region $r$, $H^r_-$;
\be
^{in}\!f_H^l=\frac{e^{i\omega t}e^{-i\omega x^*}}{\sqrt{4\pi \omega}}=\frac{e^{i\omega u}}{\sqrt{4\pi \omega}}\label{Eq:ModeDef3}
\ee
on the portion of the past horizon in region $l$, $H_-^l$.
These are positive norm modes on $I^r_-$ or $H_-$ which together form a Cauchy surface for the spacetime. These modes are associated with annihilation operators in the expansion of the field $\hat{\phi}^{(2)}$ in  Eq. (\ref{Eq:ModeExpansion1}).

\begin{figure}
\begin{tikzpicture}
\node (I)    at ( 4,0)   {};
\node (II)   at (-4,0)   {};
\node (III)  at (0, 2.5) {};
\node (IV)   at (0,-2.5) {};
\node (V)   at (2.,-2.) {};
\node (VI)   at (6.25,-1.75) {};
\node (VII)    at ( 4.,2.5)   {};
\node (VIII)   at (-2.,2.) {};

\node (IX)    at ( 0,4)   {};

\path 
   (IX) +(90:4)  coordinate (IXtop)
       +(-90:4) coordinate(IXbot)
       +(180:4) coordinate(IXleft)
       +(0:4)   coordinate(IXright)
       ;
\draw (IXleft) --
          node[midway, below, sloped] {}
      (IXtop) --
          node[midway, above right]    {}
          node[midway, below, sloped] {}
      (IXright) --
          node[midway, below right]    {}
          node[midway, above, sloped] {}
      (IXbot) --
          node[midway, above, sloped] {}
      (IXleft) -- cycle;

\path  
  (II) +(90:4)  coordinate[label=90:]  (IItop)
       +(0:4)   coordinate                  (IIright)
       ;
\draw
      (IItop) --
          node[midway, below, sloped] {$H_-$}
      (IIright)-- cycle;

\path 
   (I) +(90:4)  coordinate[label=90:$\quad i_+$]  (Itop)
       +(-90:4) coordinate[label=-90:$i_-$]  (Ibot)
       +(180:4) coordinate (Ileft)
       +(0:4)   coordinate[label=0:$i_0$] (Iright)
       ;
\draw (Ileft) --
          node[midway, below, sloped] {}
      (Itop) --
          node[midway, above right]    {}
          node[midway, below, sloped] {}
      (Iright) --
          node[midway, below right]   {$I_-^r$}
          node[midway, above, sloped] {}
      (Ibot) --
          node[midway, above, sloped] {}
      (Ileft) -- cycle;
\draw  (Ileft) -- (Itop) -- (Iright) -- (Ibot) -- (Ileft) -- cycle;

{\color{blue}

\path[->]
 (V) +(45:0) coordinate(IVtop)
	+(45:2) coordinate(IVbot)
	+(45:4) coordinate[label=45:$\quad R_I^r$](IVbotII)
	+(45:6.25) coordinate(IVRight);
	
\draw[->] (IVbotII)->(IVRight);

\path[->]	
	(I) +(135:1) coordinate[label=180:$ T_I^r\quad$](IVbotIII)
		+(135:4) coordinate[](IVbotIV);
 \draw[->] (IVbotIII)->(IVbotIV);

 \path 
 (IX) +(45:1)  coordinate[label=0:$\quad R_I^l$] (IXtopmid)
   (IX) +(45:3.5)  coordinate (IXtopouter)
   (IX) +(135:1)  coordinate[label=180:$ T_I^l\  \ $] (IXtopmidleft)
   (IX) +(135:3.5)  coordinate (IXtopleftouter);

\draw[->] (IXtopmid)  ->(IXtopouter);
\draw[->] (IXtopmidleft)  ->(IXtopleftouter);

\path[->]
 (III) +(-38:1.5) coordinate(IIIBot)
 (III) +(0:.25) coordinate(IIITop)
;

\path[->]
 (VI) +(45:0) coordinate[](IVOut)
	+(135:1.5) coordinate(IVIn);
\draw[->] (IVOut) ->(IVIn);

\path[->]
 (VI) +(90:.2) coordinate[label=90:$ ^{in}\!f_I^r  $ ](IVOut2)
	+(135:1.5) coordinate(IVIn2);
	
\path[->]
 (VIII) +(45:0) coordinate[label=45:](VIIItop)
	+(45:1.5) coordinate(VIIIbot);
}
\end{tikzpicture}
\caption{\label{Fig:Penrose2}Penrose diagram  illustrating  the scattering of an  $^{in} f_I^r  $ mode in the ${l}$ and ${r}$ regions.}
\end{figure}
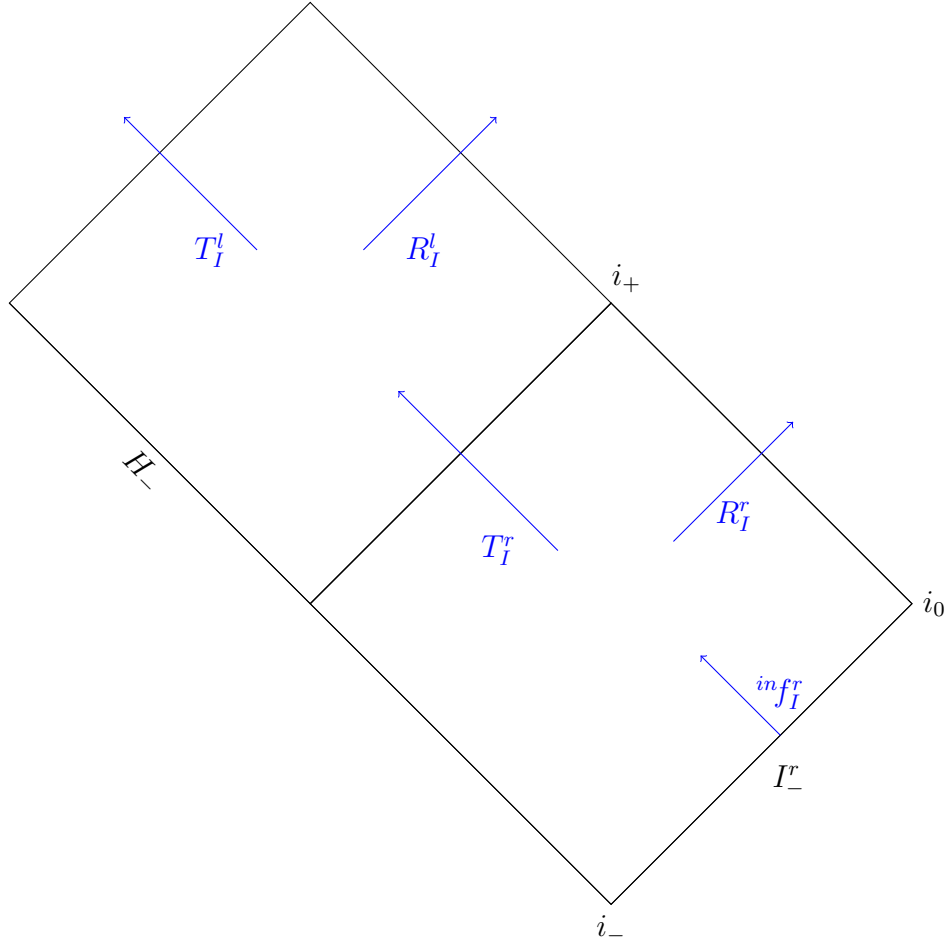

\noindent In Eqs (\ref{Eq:ModeDef1}-\ref{Eq:ModeDef3}), $u=t-x^*$ and $v=t+x^*$ are the Eddington-Finkelstein retarded and advanced null coordinates respectively.
Note the $+$ sign in the exponent of Eq. (\ref{Eq:ModeDef3}).
The conserved (Killing) energy associated with it is negative and corresponds to excitations called ``partners''.
We need to find the explicit forms of the modes throughout the spacetime. Let us begin with the $^{in}\!f_I^r$ mode whose evolution is represented schematically in Fig  \ref{Fig:Penrose2}.

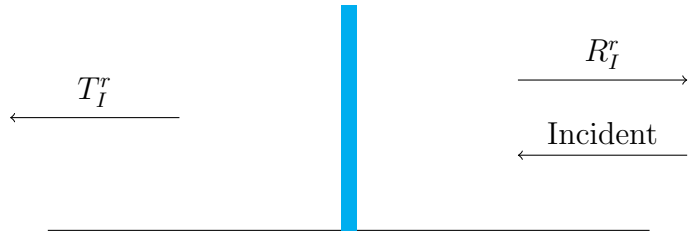
\begin{figure}
\begin{tikzpicture}
\draw[->]   (4.5,1) -- node[above] {Incident} (2.25,1);
\draw[->]   (-2.25,1.5) --node[above] {$T_I^r$} (-4.5,1.5);
\draw[->]   (2.25,2)--node[above] {$R_I^r$} (4.5,2) ;
\draw[-]   (-4,0) -- (4,0);
\fill [cyan] (-.1,0) rectangle (.1,3);
\end{tikzpicture}
\caption{\label{Fig:PlaneWaveIncident3}Illustration of a  plane wave incident onto a potential from the right, which then is partially  transmitted  to the left and also partial reflected back to the right of the barrier.}
\end{figure}

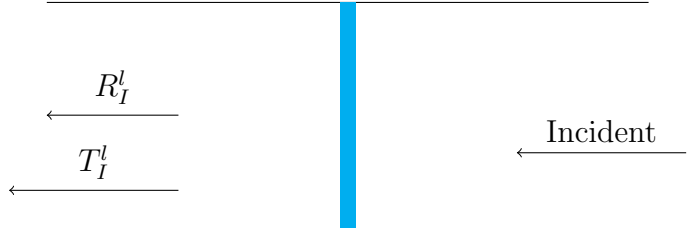
\begin{figure}
\begin{tikzpicture}
\draw[->]   (4.5,2) -- node[above] {Incident} (2.25,2);
\draw[->]   (-2.25,1.5) --node[above] {$T^l_I$} (-4.5,1.5);
\draw[->]   (-2.25,2.5)--node[above] {$R^l_I$} (-4,2.5) ;
\draw[-]   (-4,4) -- (4,4);
\fill [cyan] (-.1,1) rectangle (.1,4);
\end{tikzpicture}
\caption{\label{Fig:PlaneWaveIncident4}Illustration of a  plane wave incident onto a negative potential from the right. As this is in the interior of the BH  both the transmitted and reflected portions of the mode are forced to travel further into the BH. }
\end{figure}

The incoming $v$ mode of the form Eq. (\ref{Eq:ModeDef1}) coming from $ I^r_- $ is partially transmitted ($T_I^r$) towards the horizon as a $v$ mode and partially reflected ($R_I^r$) back to infinity $ I_+^r $  as a $u$ mode by the delta potential located at $x_r^*=0$ (see Fig.  \ref{Fig:PlaneWaveIncident3}). The transmitted part crosses the horizon, enters the black hole and is split by the second delta function potential located inside the black hole at $x_l^*=0$ (see Fig. \ref{Fig:PlaneWaveIncident4}) into a transmitted ($T_I^l$) $v$ mode and a ``reflected'' ($R_I^l$) $u$ mode both traveling inside along the flow toward left future infinity  ($I_+^l$).  Thus in the $r$ region
 \bea  ^{\rm in}f^r_I &=& \frac{e^{-i\omega t}}{\sqrt{4\pi\omega}}\left[ e^{-i \w x^*} + R^r_I e^{i \w x^*}\right]  \,, \qquad x^{*} > x^{*}_r = 0  \;, \nonumber \\
                        &=&\frac{e^{-i\omega t}}{\sqrt{4\pi\omega}} T^r_I e^{-i \w x^*}  \,, \qquad x^{*} < x^{*}_r = 0  \;, \label{chi-r-I-r-region} \eea
 and in the $\ell$ region
 \bea  ^{\rm in}f^r_I &=&\frac{e^{-i\omega t}}{\sqrt{4\pi\omega}} T^r_I e^{-i \w x^*} \,, \qquad x^{*} < x^{*}_\ell = 0  \;, \nonumber \\
                        &=&  \frac{e^{-i\omega t}}{\sqrt{4\pi\omega}}\left[ T^\ell_I e^{-i \w x^*} +  R^\ell_I e^{i \w x^*} \right] \,, \qquad x^{*} > x^{*}_l= 0  \;, \label{chi-r-I-l-region} \eea

The transmission and reflection coefficients are found by matching these solutions across the delta function potentials.
  In general for a potential of the form $V=\lambda\delta(x^*)$ we require that $\chi(x^*)$ satisfies
\bea
\left. \chi\right|_-=\left. \chi\right|_+\\
\left. \chi'\right|_+-\left. \chi'\right|_-=-\lambda \left. \chi\right|_- \ ,
\eea
where $\left.\chi\right|_\pm=\lim_{x^*\to 0^\pm}\chi$, and $\chi^\prime$ represents the derivative with respect to $x^*$.
The results for $\chi^r_I$ are
\bes \bea
T_I^r&=&\frac{\frac{2i\omega}{V_r}}{\frac{2i\omega}{V_r}-1}\ , \\
R_I^r&=&\frac{1}{\frac{2i\omega}{V_r}-1}\ ,  \\
R_I^l&=&\frac{{V_l}}{{2i\omega}}T_I^r\ , \\
T_I^l&=&\left(1-\frac{V_l}{2i\omega}\right)T_I^r\ . \eea \ees
These satisfy the relations
\bes \bea |R_I^r|^2 +|T_I^r|^2 &=& 1, \\
|T_I^l|^2 -|R_I^l|^2  +|R_I^r|^2 &=& 1\label{Eq:modifiedUnitarity1}\ .\eea \ees
The negative sign in front of the $R_I^l$ term in~(\ref{Eq:modifiedUnitarity1}) comes from the fact that the ``reflected'' modes $R_I^le^{-i\omega u}$ inside the BH have a negative norm (see Eq.~\eqref{sc}).

The asymptotic form of the $ ^{in}\!f_I^r$ mode as $x\to +\infty$ is
\be\label{ciao}
^{in}\!f_I^r=\frac{e^{-i\omega t}}{\sqrt{4\pi \omega}}\left[ e^{-i\omega x^*}+R_I^re^{i\omega x^*}\right]
\ee
 and
 \be
^{in}\!f_I^l=\frac{e^{-i\omega t}}{\sqrt{4\pi \omega}}\left[ T_I^le^{-i\omega x^*}+R_I^le^{i\omega x^*}\right]\label{Eq:InMode1c}
\ee
for $x\to -\infty$.

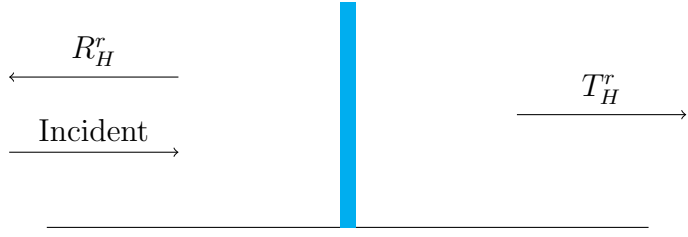
\begin{figure}
\begin{tikzpicture}
\draw[->]   (-4.5,1) -- node[above] {Incident} (-2.25,1);
\draw[->]   (-2.25,2) --node[above] {$R^r_H$} (-4.5,2);
\draw[->]   (2.25,1.5)--node[above] {$T^r_H$} (4.5,1.5) ;
\draw[-]   (-4,0) -- (4,0);
\fill [cyan] (-.1,0) rectangle (.1,3);
\end{tikzpicture}
\caption{\label{Fig:PlaneWaveIncident5}Illustration of a  plane wave incident onto a potential from the left, which then is partially  transmitted  to the right and also partial reflected back to the left of the barrier.  The reflected portion then travels into the interior of the BH where it encounters the potential in the interior. There the ``reflected'' and ``transmitted'' portions travel away from the potential to the left, see Fig. \ref{Fig:PlaneWaveIncidentb}. }
\end{figure}

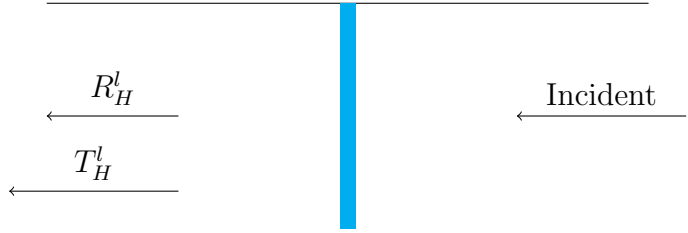
\begin{figure}
\begin{tikzpicture}
\draw[->]   (4.5,2.5) -- node[above] {Incident} (2.25,2.5);
\draw[->]   (-2.25,1.5) --node[above] {$T^l_H$} (-4.5,1.5);
\draw[->]   (-2.25,2.5)--node[above] {$R^l_H$} (-4,2.5) ;
\draw[-]   (-4,4) -- (4,4);
\fill [cyan] (-.1,1) rectangle (.1,4);
\end{tikzpicture}
\caption{\label{Fig:PlaneWaveIncidentb} Scattering inside the horizon of the mode $^{in}f_H^r$.
}
\end{figure}

Following the same procedure for the $^{in}\!f_H^r$ modes coming  out from the part of the past horizon in the $r$ region (see Fig \ref{Fig:PlaneWaveIncident5}),
we have
\be
T_H^r=\frac{1}{1-\frac{V_r}{2i\omega}}\ ,
\ee
\be
R_H^r=\frac{\frac{V_r}{2i\omega}}{1-\frac{V_r}{2i\omega}}\ ,
\ee
satisfying $|R_H^r|^2 +|T_H^r|^2 = 1$. Similarly, the ingoing $R_H^r$ part gets scattered by the $\delta$ potential inside the horizon (as shown in  Fig. \ref{Fig:PlaneWaveIncidentb})
\be
T_H^l=\left(1-\frac{V_l}{2i\omega}\right)R_H^r \;,
\ee
\be
R_H^l=\frac{{V_l}}{{2i\omega}}R_H^r \;,
\ee
again with $|T_H^l|^2 -|R_H^l|^2  +|T_H^r|^2= 1$
leading to the asymptotic form
\be \label{ciao2}
^{in}\!f_H^r=\frac{e^{-i\omega t}}{\sqrt{4\pi \omega}}T_H^re^{i\omega x^*}
\ee
 for $x\to +\infty $ and
 \be
^{in}\!f_H^r=\frac{e^{-i\omega t}}{\sqrt{4\pi \omega}}\left[ R_H^le^{-i\omega x^*}+T_H^le^{i\omega x^*}\right]\label{Eq:InMode2c}
\ee
 for $x\to -\infty $.

\begin{figure}
\begin{tikzpicture}
\draw[->]   (4.5,2.5) -- node[above] {Incident} (2.25,2.5);
\draw[->]   (-2.25,2.25) --node[above] {$\tilde T^l_H$} (-4.5,2.25);
\draw[->]   (-2.25,3.15)--node[above] {$\tilde R^l_H$} (-4,3.15) ;
\draw[-]   (-4,4) -- (4,4);
\fill [cyan] (-.1,1) rectangle (.1,4);
\end{tikzpicture}
\caption{\label{Fig:PlaneWaveIncidenta} Scattering of the mode $^{in}f_H^l$.
}
\end{figure}
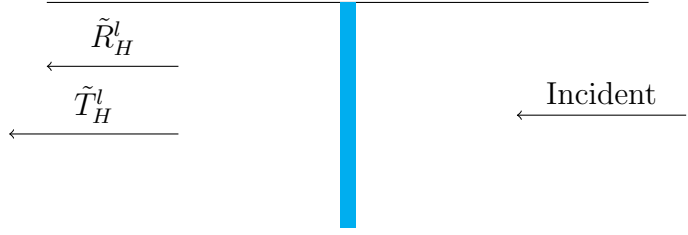

Finally, for the modes $^{in}\!f_H^l$ coming from the part of the past horizon in region $l$,
 see Fig. \ref{Fig:PlaneWaveIncidenta}, the effective transmission and reflection coefficients are
\bes \bea
\tilde{T}_H^l&=&1-\frac{V_l}{2i\omega}\ , \\
\tilde{R}_H^l&=&\frac{{V_l}}{{2i\omega}}\ ,
\eea \ees
satisfying $|\tilde{T}_H^l|^2-|\tilde{R}_H^l|^2  = 1$.
The asymptotic ($x\to -\infty$ )  form  of $^{in}\!f_H^l$ is
 \be
^{in}\!f_H^l=\frac{e^{i\omega t}}{\sqrt{4\pi \omega}}\left[ \tilde{T}_H^le^{-i\omega x^*}+\tilde{R}_H^le^{i\omega x^*}\right]\label{Eq:InMode3c}\ .
\ee
Having defined the ``in'' basis, the field operator ${\hat{\phi}^{(2)}}$ can be expanded as
\be
\hat{\phi}^{(2)}=\int d\omega \left[ _r^{in}\hat{a}_I (^{in}\!f_I^r) +\ _r^{in}\hat{a}_H (^{in}\!f_H^r)+\ _l^{in}\hat{a}_H (^{in}\!f_H^l)+ \text{h.c.}\right]
\ee
where the $\hat{a}$'s are the annihilation operators for the respective modes.

\begin{figure}
\begin{tikzpicture}
\node (I)    at ( 4,0)   {r};
\node (II)   at (-4,0)   {};
\node (III)  at (0, 2.5) {};
\node (IV)   at (0,-2.5) {};
\node (V)   at (2.,-2.) {};
\node (VI)   at (6.25,-1.75) {};
\node (VII)    at ( 4.,2.5)   {};
\node (VIII)   at (-2.,2.) {};

\node (IX)    at ( 0,4)   {l};

\path 
   (IX) +(90:4)  coordinate (IXtop)
       +(-90:4) coordinate(IXbot)
       +(180:4) coordinate(IXleft)
       +(0:4)   coordinate(IXright)
       ;
\draw (IXleft) --
          node[midway, below, sloped] {}
      (IXtop) --
          node[midway, above right]    {}
          node[midway, below, sloped] {}
      (IXright) --
          node[midway, below right]    {}
          node[midway, above, sloped] {}
      (IXbot) --
          node[midway, above, sloped] {}
      (IXleft) -- cycle;

\path  
  (II) +(90:4)  coordinate[label=90:]  (IItop)
       +(0:4)   coordinate                  (IIright)
       ;
\draw
      (IItop) --
          node[midway, below, sloped] {$H_-$}
      (IIright)-- cycle;

\path 
   (I) +(90:4)  coordinate[label=90:\quad $i_+$]  (Itop)
       +(-90:4) coordinate[label=-90:$i_-$]  (Ibot)
       +(180:4) coordinate (Ileft)
       +(0:4)   coordinate[label=0:$i_0$] (Iright)
       ;
\draw (Ileft) --
          node[midway, below, sloped] {}
      (Itop) --
          node[midway, above right]    {}
          node[midway, below, sloped] {}
      (Iright) --
          node[midway, below right]   {$I_-^r$}
          node[midway, above, sloped] {}
      (Ibot) --
          node[midway, above, sloped] {}
      (Ileft) -- cycle;
\draw  (Ileft) -- (Itop) -- (Iright) -- (Ibot) -- (Ileft) -- cycle;

{\color{blue}

\path[->]
 (V) +(45:0) coordinate(IVtop)
	+(45:2) coordinate(IVbot)
	+(45:4) coordinate[label=-45:$\  ^{out}\! f_u^r $](IVbotII)
	+(45:6.25) coordinate(IVRight);
	
\draw[->] (IVbotII)->(IVRight);

\path[->]	
	(I) +(135:1) coordinate(IVbotIII)
		+(135:4) coordinate[](IVbotIV);

 \path 
 (IX) +(45:1)  coordinate[label=0:$\ \ ^{out}\! f_u^l$ ] (IXtopmid)
   (IX) +(45:3.5)  coordinate (IXtopouter)
   (IX) +(135:1)  coordinate[label=180:$^{out}\! f_v^l \ \ $] (IXtopmidleft)
   (IX) +(135:3.5)  coordinate (IXtopleftouter);

\draw[->] (IXtopmid)  ->(IXtopouter);
\draw[->] (IXtopmidleft)  ->(IXtopleftouter);

\path[->]
 (III) +(-38:1.5) coordinate(IIIBot)
 (III) +(0:.25) coordinate(IIITop)
;

\path[->]
 (VI) +(45:0) coordinate[](IVOut)
	+(135:1.5) coordinate(IVIn);

\path[->]
 (VIII) +(45:0) coordinate[label=45:](VIIItop)
	+(45:1.5) coordinate(VIIIbot);
}
\end{tikzpicture}
\caption{\label{Fig:Penrose3}Penrose diagram  illustrating  the modes forming the ``out'' basis.
}
\end{figure}
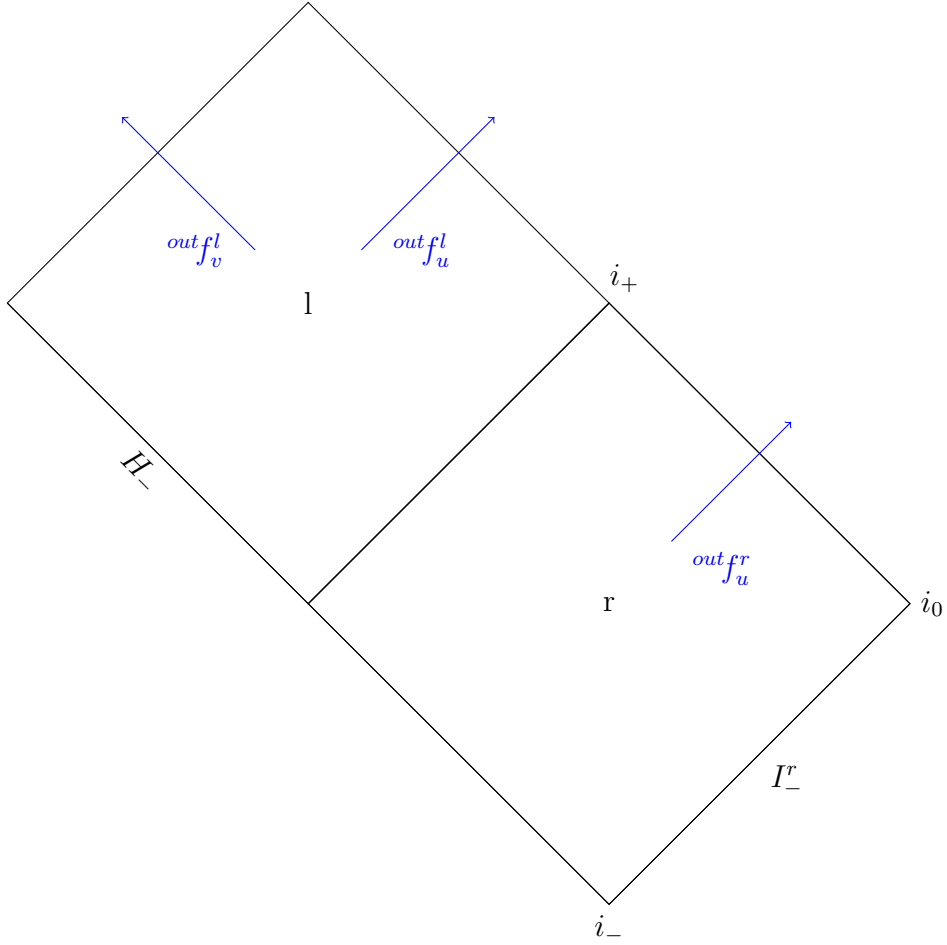

Alternatively one can construct another basis called the ``out'' basis formed by modes
having the asymptotic form

\be
^{out}\!f_u^r=\frac{e^{-i\omega t}}{\sqrt{4\pi \omega}}e^{i\omega x^*}=\frac{e^{-i\omega u}}{\sqrt{4\pi \omega}}\label{Eq:OutMode1}
\ee
 for $x\to +\infty$ and
 \be
^{out}\!f_u^l=\frac{e^{i\omega t}}{\sqrt{4\pi \omega}}e^{-i\omega x^*}=\frac{e^{i\omega u}}{\sqrt{4\pi \omega}}\ , \label{Eq:OutMode2}
\ee
 \be
^{out}\!f_v^l=\frac{e^{-i\omega t}}{\sqrt{4\pi \omega}}e^{-i\omega x^*}=\frac{e^{-i\omega v}}{\sqrt{4\pi \omega}}\label{Eq:OutMode3}
\ee
for $x\to -\infty $. These modes are represented in the Penrose diagram in Fig.\ref{Fig:Penrose3}.

Proceeding in the same manner we can construct the $^{out}\!f$ modes throughout the spacetime and then obtain the following expressions of the field operator
\be
\hat{\phi}^{(2)}=\int d\omega \left[ _r^{out}\hat{a}_u\ (^{out}\!f_u^r) +\ _l^{out}\hat{a}_u\ (^{out}\!f_u^l)+\ _r^{out}\hat{a}_v\ (^{out}\!f_v^r)+ \text{h.c.}\right]
\ee
where the $^{out}\hat{a}$'s are the associated annihilation operators.
The ``in'' and ``out'' basis are related by a Bogoliubov transformation. Looking at the asymptotic form of the in modes Eqs. (\ref{ciao}, \ref{Eq:InMode1c}, \ref{ciao2}, \ref{Eq:InMode2c}) and (\ref{Eq:InMode3c}),  one can rewrite the modes on $I_+$ as follows
\be
^{in}\!f_I^r=R_I^r \  ^{out}\!f_u^r+T_I^l \ ^{out}\!f_v^l + R_I^l \  ^{out}\!f_u^{l*},\label{Eq:IntoOut1}
\ee
 \be
^{in}\!f_H^r=T_H^r \ ^{out}\!f_u^r+T_H^l \ ^{out}\!f_v^l + R_H^l \  ^{out}\!f_u^{l*} ,\label{Eq:IntoOut2}
\ee
 and
 \be
^{in}\!f_H^l=  \tilde{R}_H^l \ ^{out}\!f_v^{l*} + \tilde{T}_H^l \  ^{out}\!f_u^{l}.\label{Eq:IntoOut3}
\ee
Note that there is no contribution to $^{in}\!f_H^l$ from the $^{out}\!f_u^r$ modes.
Using the scattering S-matrix formalism we can write the relation between the two basis as
\be
\begin{pmatrix}^{in}\!f_I^r\\
^{in}\!f_H^r \\
^{in}\!f_H^{l*} \end{pmatrix}=S^T\begin{pmatrix}^{out}\!f_u^r\\
^{out}\!f_v^l \\
^{out}\!f_u^{l*} \end{pmatrix}\label{Eq:TransposeSMatrix}
\ee
where
\be
S^T=\begin{pmatrix}S_{u_r,v_r}&S_{v_l,v_r}&S_{u_l,v_r}\\
S_{u_r,u_r}&S_{v_l,u_r}&S_{u_l,u_r}\\
0&S_{v_l,u_l}&S_{u_l,u_l} \end{pmatrix}
\ee
is the transpose of the scattering matrix $S$.

The notation used is borrowed from Ref \cite{libro}  and is quite intuitive. For example $S_{u_r,v_r}$ indicates an incoming $v$ mode from $r$ leading to an outgoing $u$ mode in $r$. The corresponding Bogoliubov transformation for the annihilation operators of the two bases is
\be
\begin{pmatrix}^{out}\hat{a}_u^r\\
^{out}\hat{a}_v^l \\
^{out}\hat{a}_u^{l\dagger} \end{pmatrix}=S\begin{pmatrix}^{in}\hat{a}_I^r\\
^{in}\hat{a}_H^r \\
^{in}\hat{a}_H^{l\dagger} \end{pmatrix}\label{Eq:SOutotInMatrix1}
\ee
where
\be
S=\begin{pmatrix}S_{u_r,v_r}&S_{u_r,u_r}&0\\
S_{v_l,v_r}&S_{v_l,u_r}&S_{v_l,u_l}\\
S_{u_l,v_r}&S_{u_l,u_r}&S_{u_l,u_l} \end{pmatrix}
\ee
For the two delta functions potential the $S$-matrix elements can be found by inspection of
Eqs. (\ref{Eq:IntoOut1}\ - \ref{Eq:IntoOut3})
resulting in
\bea
S_{u_r,v_r}=R_I^r \ , \quad &S_{u_r,u_r}=T_H^r\ , \\
S_{v_l,v_r}=T_I^l \ , \quad &S_{v_l,u_r}=T_H^l \ , \\
S_{u_l,v_r}=R_I^l \ ,\quad &S_{u_l,u_r}=R_H^l\ , \\
S_{v_l,u_l}=\tilde{R}_H^{l*}\ , \quad &S_{u_l,u_l}=\tilde{T}_H^{l*}\ .
\eea
We are interested in the numbers of outgoing particles in the various channels, namely
\be
\left<_r^{out}\hat{a}^\dagger_u\   _r^{out}\hat{a}_u\right>, \ \left<_l^{out}\hat{a}^\dagger_v\   _l^{out}\hat{a}_v\right> \ \text{and} \ \left<_l^{out}\hat{a}^\dagger_u\   _l^{out}\hat{a}_u\right>.\label{Eq:ParticleNumbers1}
\ee
To perform this calculation we have first to specify the quantum state of the $\hat{\phi}^{(2)}$ operator in which the expectation values in Eq. (\ref{Eq:ParticleNumbers1}) have to be taken.
The ``in'' modes used in the expression of the field operator $\hat{\phi}^{(2)}$ have a temporal part  $e^{\pm i\omega t}$. These are the eigenfunctions of the Killing vector $\frac{\partial}{\partial t}$ associated with the stationarity of the metric and are positive or negative (Killing) energy modes with respect to Schwarzschild time $t$.  The quantum state associated with this expansion is annihilated by all the $^{in}\hat{a}$ operators and is called the Boulware vacuum \cite{boulware}, i.e.,
\bea
&^{in}\hat{a}_I^r\left|B\right>=0\ , \nonumber\\
&^{in}\hat{a}_H^r\left|B\right>=0\ ,  \nonumber\\
&^{in}\hat{a}_H^l\left|B\right>=0\ \
\eea
for all values of $\omega$.
This is the most ``natural'' quantum state one can define on the extended manifold described by the Penrose diagram of Fig \ref{Fig:Penrose1}. Physically $\left|B\right>$ describes a state in which there are no incoming particles either from past right  infinity $I^r_-$ or from the past horizon $H_-$. Although ``natural'', this does not correctly describe the quantum state of the field $\hat{\phi}^{(2)}$ if the BH is formed by a dynamic gravitational collapse. The collapse in fact induces the conversion of quantum vacuum fluctuations to real on shell particles, the so called Hawking radiation \cite{hawking}. The state which correctly describes this process, at least at late times, is called the Unruh vacuum $\left|U\right>$ \cite{unruh76}. The difference between the two states can be schematically summarized as follows. For the Unruh vacuum the modes coming out from the past horizon are chosen to be positive and negative frequency, not with respect to the Schwarzschild time $
 t$, but with respect to Kruskal time.  Thus instead of the mode $^{in}\!f_u^r$ and $^{in}\!f_u^l$, the modes are chosen as
\bea
f_H^K=\frac{e^{-i\omega_K U}}{\sqrt{4\pi\omega}}\ , \\
U=\mp\frac{e^{-\kappa u}}{\kappa}
\eea
where the $-$ and $+$ refer to the $r$ and $l$ regions respectively, and $\kappa$ is the surface gravity of the BH horizon, which for our metric is
\be
\left.\kappa=\frac{1}{2c}\frac{d}{dx}\left(c^2-v^2\right)\right|_{x=0}.
\ee
The modes coming from past null infinity for the Unruh vacuum are chosen as $^{in}\!f_I^r$. The field can then be expanded in terms of a complete set of these modes
\be
\hat{\phi}^{(2)}=\int d\omega_k \left[(\hat{a}_{\omega_K}\ f_H^K +\hat{a}_{\omega_K}^\dagger\ f_H^{K*}\right) +\int d\omega \left[^{in}\hat{a}_I^r\ (^{in}\!f_I^r)+\ ^{in}\hat{a}_I^{r\dagger}\ (^{in}\!f_I^{r*})\right].
\ee
The Unruh state is therefore defined as
\bea
\hat{a}_{\omega_K}\left|U\right>=0\ ,  \nonumber\\
^{in}\hat{a}_I^r\left|U\right>=0 \ ,
\eea
for every $\omega$ and $\omega_K$.
The relation between the two sets of operators is given by the following Bogoliubov transformations
\bea
^{in}\hat{a}_H^r=\int d\omega_k \left[\alpha^r_{\omega_K \omega}\hat{a}_{\omega_K} +\beta^{r*}_{\omega_K \omega}\hat{a}_{\omega_K}^\dagger\right]\ , \nonumber \\
^{in}\hat{a}_H^l=\int d\omega_k \left[\alpha^l_{\omega_K \omega}\hat{a}_{\omega_K} +\beta^{l*}_{\omega_K \omega}\hat{a}_{\omega_K}^\dagger\right]\ , \label{Eq:IntoKruskalOperators}
\eea
where the Bogoliubov coefficients are given by (see for example \cite{paper2013})
\bea
\alpha^r_{\omega_K \omega}=\frac{1}{2\pi\kappa}\sqrt{\frac{\omega}{\omega_K}}(-i\omega_K)^{\frac{i\omega}{\kappa}}\Gamma \left(\frac{-i\omega}{\kappa}\right)\ , \nonumber \\
\beta^r_{\omega_K \omega}=\frac{1}{2\pi\kappa}\sqrt{\frac{\omega}{\omega_K}}(-i\omega_K)^{-\frac{i\omega}{\kappa}}\Gamma \left(\frac{i\omega}{\kappa}\right)\ , \nonumber \\
\alpha^l_{\omega_K \omega}=\frac{1}{2\pi\kappa}\sqrt{\frac{\omega}{\omega_K}}(i\omega_K)^{-\frac{i\omega}{\kappa}}\Gamma \left(\frac{i\omega}{\kappa}\right)\ , \nonumber \\
\beta^l_{\omega_K \omega}=\frac{1}{2\pi\kappa}\sqrt{\frac{\omega}{\omega_K}}(i\omega_K)^{\frac{i\omega}{\kappa}}\Gamma \left(\frac{-i\omega}{\kappa}\right).
\eea

\begin{figure}[h]
\includegraphics[width=6.cm]{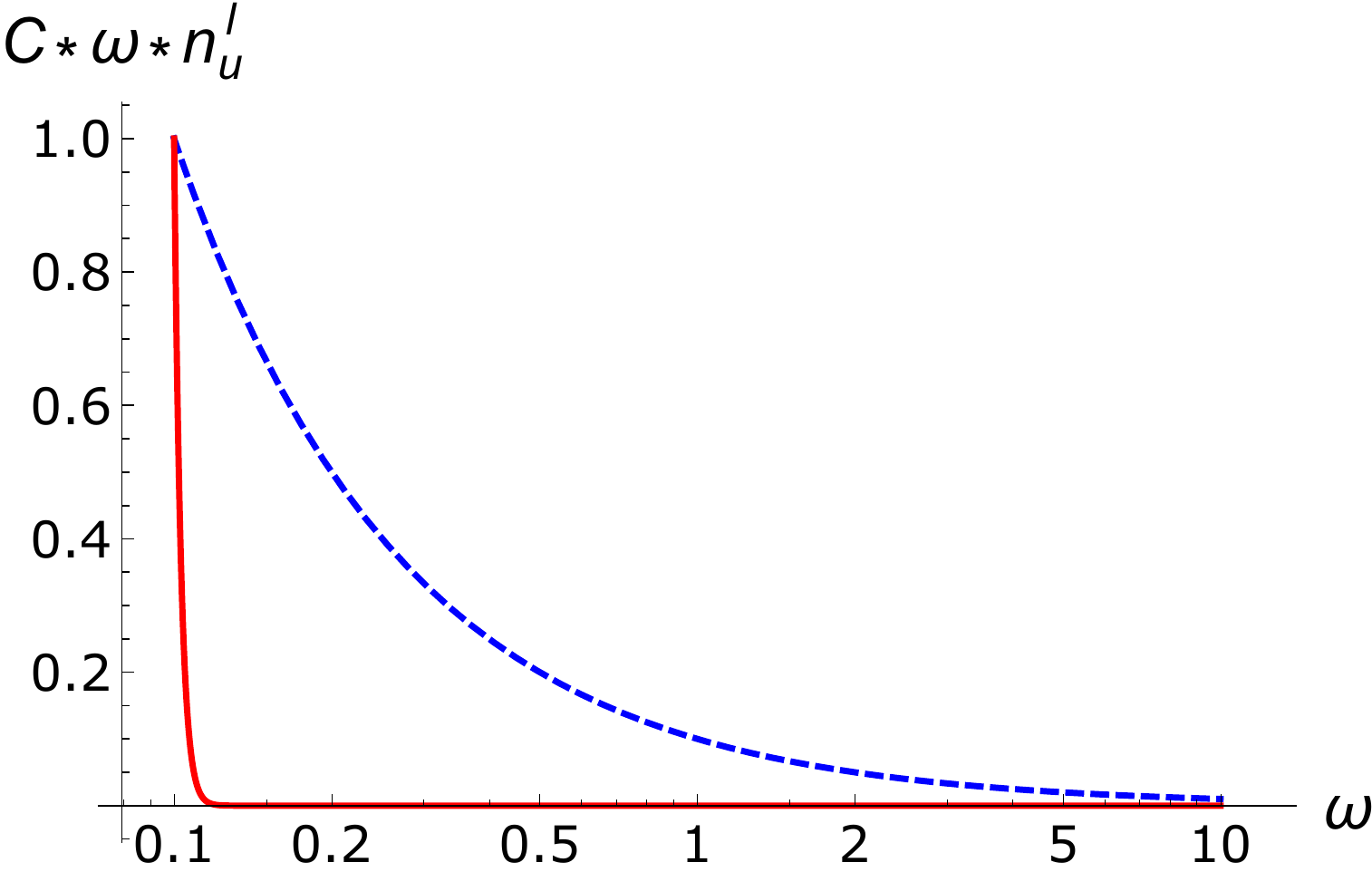}
\caption{\label{Fig:NoPotentialComparison} Plots for $\omega>>\kappa$  with $-V_l=V_r/10=\kappa/100$ (Blue, Dashed) and $V_l=V_r=0$ (Red, Solid). The quantity $C\ \omega \ n^l_u$, where $C$ is a different scaling constant so that $C\ \omega \ n^l_u=1$ for ${\omega = 10^{-1}}$.
   }
\end{figure}

Using the Bogoliubov transformations, Eq. (\ref{Eq:SOutotInMatrix1}) and  Eq. (\ref{Eq:IntoKruskalOperators}), we obtain
\be
n_u^r\equiv\left<U\right|_r^{out}\hat{a}^\dagger_u\   _r^{out}\hat{a}_u\left|U\right>=\int d\omega_K\left|S_{u_r,u_r}\right|^2\left|\beta_{\omega_K \omega}^r\right|^2\ ,
\ee
\bea
n_v^l\equiv\left<U\right|_l^{out}\hat{a}^\dagger_v\   _l^{out}\hat{a}_v\left|U\right>=\int d\omega_K\left[\left|S_{v_l,u_r}\right|^2\left|\beta_{\omega_K \omega}^r\right|^2+S_{v_l,u_r}^* \beta_{\omega_K \omega}^rS_{v_l,u_l}^*\alpha_{\omega_K \omega}^{l*}\right.\nonumber \\ \left.+S_{v_l,u_l}\alpha_{\omega_K \omega}^lS_{v_l,u_r}\beta_{\omega_K \omega}^{r*}+\left|S_{v_l,u_l}\right|^2\left|\alpha_{\omega_K \omega}^l\right|^2 \right]\ ,
\eea
\bea
n_u^l\equiv\left<U\right|_l^{out}\hat{a}^\dagger_u\   _l^{out}\hat{a}_u\left|U\right>=\int d\omega_K\left[\left|S_{u_l,u_r}\right|^2\left|\alpha_{\omega_K \omega}^r\right|^2+S_{u_l,u_r}^*\alpha_{\omega_K \omega}^{r*}S_{u_l,u_l} \beta_{\omega_K \omega}^l\right.\nonumber \\ \left.+S_{u_l,u_r}\alpha_{\omega_K \omega}^rS_{u_l,u_l}^*\beta_{\omega_K \omega}^{l*}+\left| S_{u_l,u_l}\right|\left|\beta_{\omega_K \omega}^l\right|^2 \right]+\left|S_{u_l, v_r}\right|^2 \ .
\eea
One can see the combined effect of the near horizon mixing  (the $\alpha$ and $\beta$) encoded in the Bogoliubov transformation (\ref{Eq:IntoKruskalOperators}), which engenders Hawking thermal radiation, and the scattering caused by the potential (the S matrix element).
After some calculation we obtain
\bea
n_u^r&=&\frac{4\omega^2}{4\omega^2+V_R^2}\frac{\delta (0)}{e^{\frac{2\pi\omega}{\kappa}}-1}\ ,  \label{cu}\\
n_v^l&=&\frac{1}{4\omega^2}\left|V_r\frac{2i\omega -V_l}{2i\omega -V_r}-V_le^{\frac{\pi\omega}{\kappa}}\right|^2\frac{\delta (0)}{e^{\frac{2\pi\omega}{\kappa}}-1}\ ,  \label{cu1}\\
n_u^l&=&\frac{1}{4\omega^2}\left|e^{{\frac{\pi\omega}{\kappa}}}\frac{V_rV_l}{2i\omega +V_r}+(2i\omega-V_l)\right|^2\frac{\delta (0)}{e^{\frac{2\pi\omega}{\kappa}}-1}+\frac{V_l^2}{4\omega^2+V_r^2}\ . \label{cu2}
\eea
The latter expression represents the numbers of the negative energy excitations  created inside the BH.
Here we see the usual problem of the normalization of plane waves leading to the $\delta(0)$. Using wave packets we can set $\delta (0)\to 1$ and verify that $n_u^r+n_v^l=n_u^l$. Thus the number of positive energy excitations created equals the  number of negative ones as energy conservation requires. One notices immediately the striking difference between the emission in the exterior region compared to that of the interior region.  In the exterior region the scattering is the standard one, $n_u^r$ describes, as expected, a thermal emission at the temperature ${T_H=\frac{\hbar \kappa}{2\pi k_B}}$ modulated by the gray body factor $\frac{4 \omega^2}{4\omega^2+V_r^2}$ which regulates the infrared divergence associated with the Planckian distribution.  The gray body factor goes to one for $\omega \gg V_r$. In the interior region the scattering is anomalous resulting in particle production; as a result we see that both $n_u^l$ and $n_v^l$ do not decay exponentially for large $\omega$ but as a power law as seen in Fig. \ref{Fig:NoPotentialComparison}. The emission in the interior is not thermal. It is infrared divergent, i.e, the spectrum is dominated by soft phonons.

\begin{figure}
\includegraphics[width=6.cm]{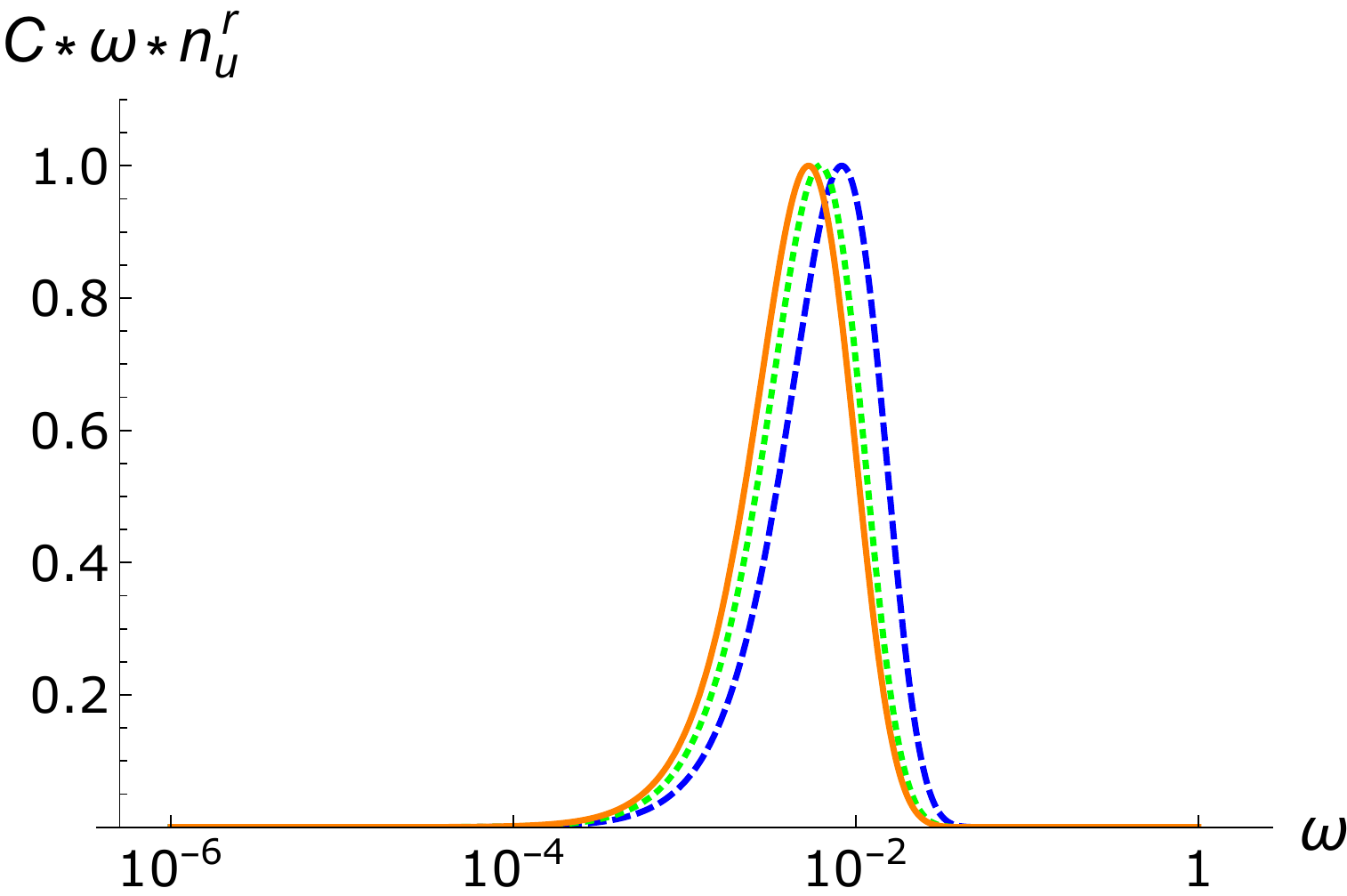}
\includegraphics[width=6.cm]{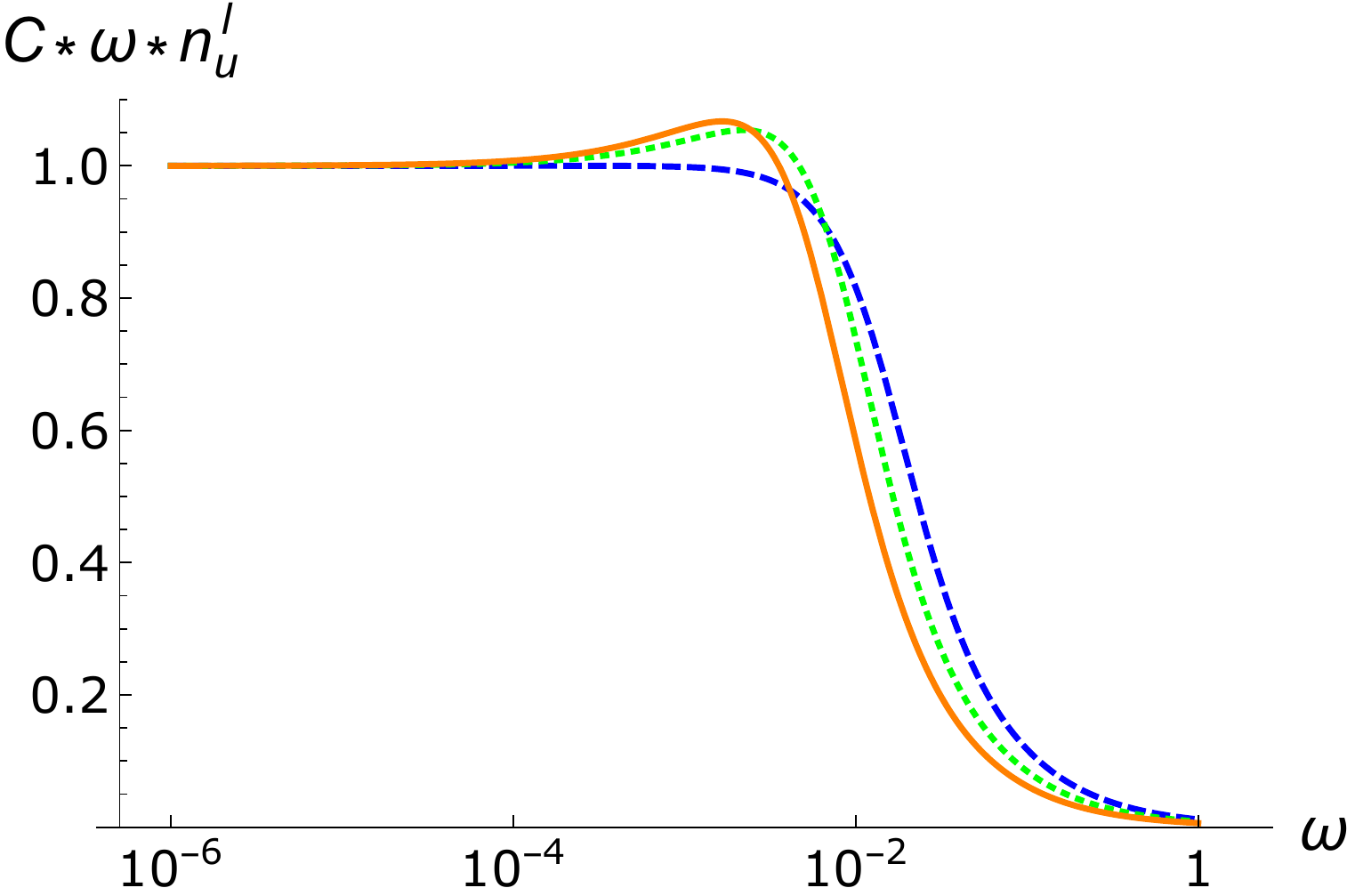}
\caption{\label{Fig:nulnvlComparison} Plots for $-V_l=V_r/10=\kappa/100$ (Blue, Dashed), $-V_l=V_r/100=\kappa/100$ (Green, Dotted) and $-V_l=3V_r/200=\kappa/100$ (Orange, Solid). Left: The quantity $C\ \omega \ n^r_u$ shows how the thermal nature of the exterior modes is modified by the gray-body factor. Here $C$ is a  scaling factor so that maximum value of $C\ \omega \ n^r_u=1$.  Right: The quantity $C\ \omega \ n^l_u$, where $C$ is a different scaling constant so that $C\ \omega \ n^l_u=1$ for ${\omega = 10^{-6}}$. An unexpected peak in $C\ \omega \ n^l_u$ is visible, and exists in all 3 cases shown. The qualitative behavior of $n^l_v$ is identical to $n^l_u$ and so is not depicted here.
   }
\end{figure}

The low frequency behaviors of $n^r_u$ and $n^l_u$ for various values of $V_r>0$ and $V_l<0$ are shown in Fig .\ref{Fig:nulnvlComparison}.  The qualitative behaviors of $n^\ell_v$ for the same cases are identical to that of $n^l_u$ and thus are not depicted.   In addition to the non thermal behaviors of the high frequency modes for $n^l_u$ seen in Fig. \ref{Fig:NoPotentialComparison}, the plots in Fig.\ref{Fig:nulnvlComparison} show another nontrivial feature, a peak, that arises in the quantity $\omega n^l_u$ ( and also occurs for $\omega n^l_v$). It appears the peak is most pronounced when $|V_l|>>V_r\sim \kappa/(2\pi)$.  In this regime, the position, in $\omega$, of the peak is proportional to $V_r$ so it moves to the right on a plot of $\w n^l_u$ versus $\w$ as $V_r$ increases.  For $V_r >> \kappa/(2\pi) $  it disappears because it becomes lost in the power law decay that occurs at high frequencies.
In contrast, as $V_r$ gets smaller and moves to the left on the plot, the height of the peak decreases relative to its base, which for small values of $\w$ is the limit ${\lim_{\w \to 0}}\ \omega n^l_u$.

If for fixed $V_r$,  $|V_l|$ decreases, but is still larger than $V_r$, then the height of the peak also decreases.  However, its location stays about the same. When $V_r=|V_l|$ the peak no longer exists. This can be shown analytically by looking at the derivative of $\omega n^l_v$,

\be
\frac{d(\omega n^l_v)}{d \omega}=\frac{{V_r}^2 \left(\kappa  \left(1-e^{\frac{2 \pi  \omega }{\kappa }}\right)+2 \pi  \omega  e^{\frac{\pi  \omega }{\kappa }}\right)}{4 \kappa  \omega ^2 \left(e^{\frac{\pi  \omega }{\kappa }}+1\right)^2}.
\ee
As the denominator is positive for all $\omega$  we can just focus at the numerator. Making the substitution $\omega^\prime=\frac{\pi \omega}{\kappa}$ it can be shown that
\be
{2{{V_r}^2}{\kappa}e^{\frac{ \omega^\prime }{2 }} \left(  \omega^\prime -\sinh \left({ \omega^\prime } \right)\right)}
\ee
 which is less than zero for all $\omega>0$. Thus there is no peak  like the one seen in Fig. \ref{Fig:nulnvlComparison} in the $V_r<|V_l|$ case. The same can also be shown for $n^\ell_u$ but the expressions are more complicated.

 This peak is also present in other, more realistic, configurations for the effective potential. This will be shown elsewhere.

A final remark concerning the Boulware vacuum $\left|B\right>$.  This state is characterized by being a vacuum state at infinity ${I^r_\pm}$ (no incoming and no outgoing particles for $x\to +\infty $), that is singular however at $H_\pm$. Indeed if we calculate the number of particles created in the $r$ region one finds
\be
N_u^r\equiv\left<B\right|_r^{out}\hat{a}^\dagger_u\   _r^{out}\hat{a}_u\left|B\right>=0\ .
\ee
This is not true in the BH interior region because of the particle production that occurs there resulting in
\be
N_v^l\equiv\left<B\right|_l^{out}\hat{a}^\dagger_v\   _l^{out}\hat{a}_v\left|B\right>=\left|S_{v_l,u_l}\right|^2=\left|\tilde{R}_H^{l*}\right|^2=\frac{V_l^2}{4\omega^2}
\ee
and
\bea
N_u^l&\equiv\left<B\right|_l^{out}\hat{a}^\dagger_u\   _l^{out}\hat{a}_u\left|B\right>=\left|S_{u_l,v_l}\right|^2+\left|S_{u_l,v_l}\right|^2\nonumber \\& =\left|{R}_I^{l}\right|^2+\left|{R}_H^{l}\right|^2=\frac{V_l^2}{4\omega^2}\left[\left|{R}_H^{r}\right|^2+\left|{T}_H^{r}\right|^2\right]=\frac{V_l^2}{4\omega^2}\ .
\eea

Inside the BH $\left|B\right>$ is no longer an out vacuum state. 
Instead there is a net flux of particles (of positive and negative energy) directed towards $x\to-\infty$ with $N_v^l=N_u^l$.

\section{The massive model}
\begin{figure}[h]
\includegraphics[width=4in]{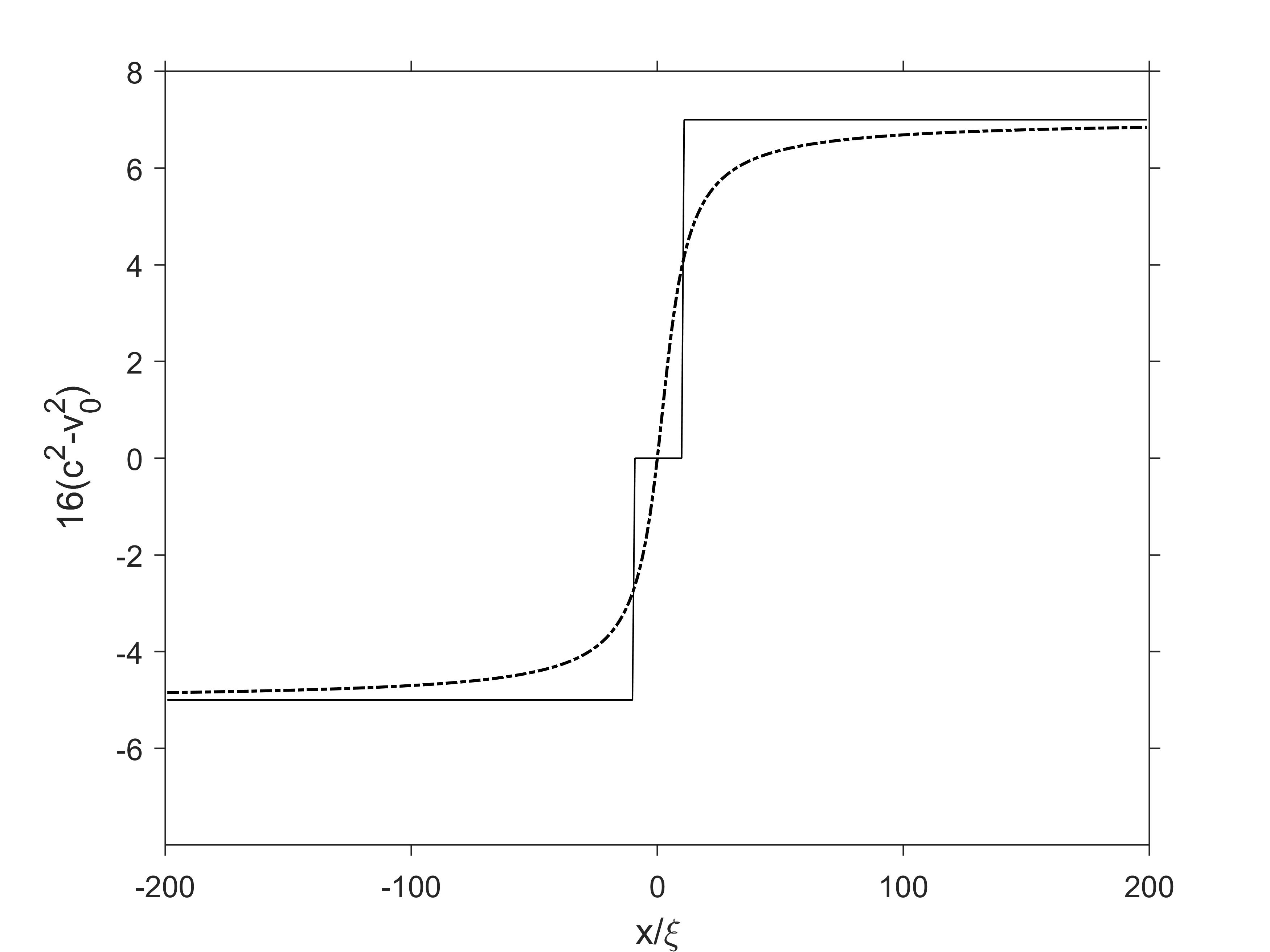}
\caption{\label{Fig:massliketerm1}  Plot from Ref  \cite{massive2} of the coefficient of $m^2$(dashed)  and the approximation to that coefficient(solid). This result is based on the sound speed profile used in ~\cite{paper2} and \cite{paper2013}.  }
\end{figure}

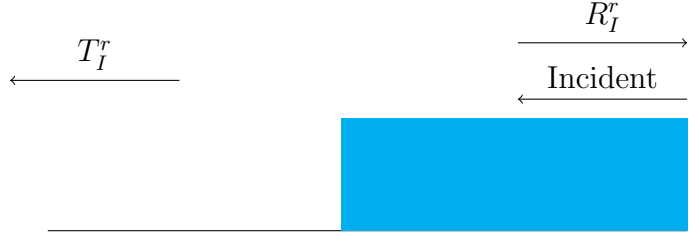
\begin{figure}
\begin{tikzpicture}
\draw[->]   (4.5,1.75) -- node[above] {Incident} (2.25,1.75);
\draw[->]   (-2.25,2) --node[above] {$T_I^r$} (-4.5,2);
\draw[->]   (2.25,2.5)--node[above] {$R_I^r$} (4.5,2.5) ;
\draw[-]   (-4,0) -- (4.5,0);
\fill [cyan] (-.1,0.005) rectangle (4.5,1.5);
\end{tikzpicture}
\caption{\label{Fig:PlaneWaveIncident6}Illustration of a  plane wave incident from $x=+\infty$,
which then is partially  transmitted  and partial reflected back.}
\end{figure}

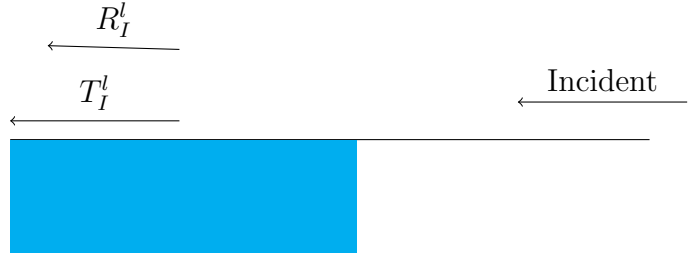
\begin{figure}
\begin{tikzpicture}
\draw[->]   (4.5,.5) -- node[above] {Incident} (2.25,.5);
\draw[->]   (-2.25,.25) --node[above] {$T^l_I$} (-4.5,.25);
\draw[->]   (-2.25,1.2)--node[above] {$R^l_I$} (-4,1.25) ;
\draw[-]   (-4.5,0) -- (4,0);
\fill [cyan] (.1,-0.005) rectangle (-4.5,-1.5);
\end{tikzpicture}
\caption{\label{Fig:PlaneWaveIncident7}Illustration of a  plane wave incident onto the negative step function potential from the right. As this is in the interior of the BH  both the transmitted and reflected portions of the mode are forced to travel further into the BH. }
\end{figure}

The second toy model we want to investigate is the one introduced in Ref \cite{massive2}, where in the field equation (\ref{Eq:WaveEquation2})  $V_{\text{eff}}$ is neglected and the mass term $k_\perp^2(c^2-v_0^2)$  is  approximated as two step functions (see Fig \ref{Fig:massliketerm1} )
\be
k_\perp^2(c^2-v_0^2)\to\begin{cases}
 \ \ m_r^2\Theta(x^*-x_{0r}^*)\ , \quad x>0\ , \\
-m_l^2\Theta(x^*-x_{0l}^*)\ , \quad x<0\ ,
\end{cases}
\ee
where $m_r^2=m^2 \left(c^2_r-v_0^2\right)$  and $m_l^2=m^2 \left(v_0^2-c^2_l\right)$. Again $c_r$ and $c_l$ are the asymptotic values of $c(x)$ as $x\to +\infty$ and $x\to -\infty$ respectively. The $-$ sign in front of $m_l^2$ comes from the fact that inside the BH $c^2<v_0^2$. We also choose $x_{0l}^*=0=x_{0r}^*$ for simplicity.
The field equation (\ref{Eq:WaveEquation2}) simplifies to
\bea
[ -\partial_t^2 +\partial_{x^*}^2 - m_r^2\Theta(x^*)]\hat{\phi}^{(2)} =0\ ,  \quad x>0\ ,  \\ \relax
[ -\partial_t^2 +\partial_{x^*}^2 +m_l^2\Theta(x^*)]\hat{\phi}^{(2)} =0\ ,  \quad x<0\ ,
\eea
Since the construction of the  ``in'' basis for this model has been performed in Ref \cite{massive2}, here we briefly sketch the basic features.
The asymptotic form of the incoming $v$ mode coming from $x\to+\infty$ is
\be
^{in}\!f_I^r=\frac{1}{\sqrt{4\pi \omega}}e^{-i\omega t}e^{-ik_rx^*}  \label{massivemodes0}
\ee
with $k_r\equiv\sqrt{\omega^2-m_r^2}$.
 This is a massive mode and it exists only if $\omega>m_r$ i.e., there is, as usual, a mass gap.  On the other hand on $H^-$ where these modes are massless
 \bea
^{in}\!f_H^r=\frac{1}{\sqrt{4\pi \omega}}e^{-i\omega u}   \\ \relax
^{in}\!f_H^l=\frac{1}{\sqrt{4\pi \omega}}e^{i\omega u} \;.
\eea
The form of these modes throughout the spacetime can be found by enforcing continuity of the spatial part $\chi$ of the modes and their derivatives at the boundaries of the step functions with the result
\bea
^{in}\!f_I^r=\frac{e^{-i\omega t}}{\sqrt{4\pi\omega}}\left[\frac{k_r-\omega}{k_r+\omega}e^{ik_rx^*}+e^{-ik_rx^*}\right]\ , \quad &\text{for } x\to +\infty\ ,  \label{massivemodes1}\\
^{in}\!f_I^r=\frac{e^{-i\omega t}}{\sqrt{4\pi k_l}}\frac{\sqrt{k_lk_r}}{k_r+\omega}\left[\frac{k_l+\omega}{2k_l}e^{ik_lx^*}+\frac{k_l-\omega}{2k_l}e^{-ik_lx^*}\right]\ , \quad &\text{for } x\to -\infty\ , \label{massivemodes2}
\eea
 where $k_l\equiv\sqrt{\omega^2+m_l^2}$. Note that unlike $k_r$, $k_l$ is real for any value of omega and $k_l\geq m_l$. These modes can be illustrated schematically in the same way as the  previous toy model of Sec \ref{Sec:DiracDeltaFunctionPotential}. The scattering of these modes in the exterior is illustrated in Fig. \ref{Fig:PlaneWaveIncident6}, while Fig. \ref{Fig:PlaneWaveIncident7} illustrates the interior scattering.

The $R_I^r$ part is the coefficient of the first exponential in Eq. (\ref{massivemodes1}), while $T_I^l$ and $R_I^l$ are the coefficients of the first and second exponentials respectively in Eq. (\ref{massivemodes2}). 

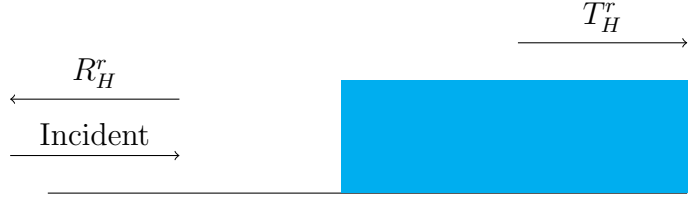
\begin{figure}
\begin{tikzpicture}
\draw[->]   (-4.5,.5) -- node[above] {Incident} (-2.25,.5);
\draw[->]   (-2.25,1.25) --node[above] {$R_H^r$} (-4.5,1.25);
\draw[->]   (2.25,2)--node[above] {$T_H^r$} (4.5,2) ;
\draw[-]   (-4,0) -- (4.5,0);
\fill [cyan] (-.1,0.005) rectangle (4.5,1.5);
\end{tikzpicture}
\caption{\label{Fig:PlaneWaveIncident8}Illustration of a  plane wave incident onto a step function potential from the left, which then is partially  transmitted  to the right and also partial reflected back to the left of the barrier.  There, the reflected portion then travels into the interior of the BH where it encounters the step function potential in the interior. There the ``reflected'' and ``transmitted'' portions travel away from the potential to the left, see Fig. \ref{Fig:PlaneWaveIncident7}. }
\end{figure}

For the $^{in}\!f_H^r$ modes one finds
\bea
^{in}\!f_H^r=\frac{e^{-i\omega t}}{\sqrt{4\pi k_r}}\frac{2\sqrt{k_r\omega}}{k_r+\omega}e^{ik_rx^*} \quad &\text{for } x\to +\infty\ , \label{massivemodes3}\\
^{in}\!f_H^r=\frac{e^{-i\omega t}}{\sqrt{4\pi k_l}}\left(\sqrt{\frac{k_l}{\omega}}\frac{\omega-k_r}{k_r+\omega}\right)\left[\frac{k_l+\omega}{2k_l}e^{-ik_lx^*}+\frac{k_l-\omega}{2k_l}e^{ik_lx^*}\right]\quad &\text{for } x\to -\infty \ . \label{massivemodes4}
\eea
Schematically the exterior scattering is described in Fig. \ref{Fig:PlaneWaveIncident8} and the inner one is similar to the one represented in Fig. \ref{Fig:PlaneWaveIncident7}.

\noindent The $T_H^r$ term is the coefficient of the first exponential in Eq. (\ref{massivemodes3}), while $T_H^l$  and $R_H^l$  are the coefficients of the first and second exponential respectively in Eq. (\ref{massivemodes4}). Note that for $\omega<m_r$ the $^{in}\!f_H^r$ mode coming from $H^r_-$ is completely reflected at $x_{0r}^*$. This is the boomerang effect as seen in Ref \cite{Rousseaux2011}.

The final set of modes in this basis are the $^{in}\!f_H^l$ modes which are
\be
^{in}\!f_H^l=\frac{e^{-i\omega t}}{\sqrt{4\pi k_l}}\sqrt{\frac{k_l}{\omega}}\left[\frac{k_l+\omega}{2k_l}e^{-ik_lx^*}+\frac{k_l-\omega}{2k_l}e^{ik_lx^*}\right]\quad \text{for } x\to -\infty\ . \label{massivemodes5}
\ee
Schematically this is the same as seen in Fig. \ref{Fig:PlaneWaveIncident7}.
 $\tilde T_H^l$ is the coefficient of first exponential in Eq. (\ref{massivemodes5}) and $\tilde{R}_H^l$ is the coefficient of the second one.

\noindent The ``out'' basis is constructed by a similar procedure to that described in the previous section starting from the asymptotic form of the modes
\be
^{out}\!f_u^r=\frac{1}{\sqrt{4\pi k_r}}e^{-i\omega t}e^{ik_rx^*}\ ,    \ee
as $x\to +\infty$, and
\bea  ^{out}\!f_u^l=\frac{1}{\sqrt{4\pi k_l}}e^{i\omega t}e^{-ik_lx^*}\ ,   \\ \relax
^{out}\!f_v^l=\frac{1}{\sqrt{4\pi k_l}}e^{-i\omega t}e^{-ik_lx^*}\ ,
\eea
for $x\to -\infty$.
From Eq (\ref{massivemodes1}\ -\ \ref{massivemodes5}) we can express the ``in'' modes in terms of the ``out'' modes
as
\bea
^{in}\!f_I^r=\frac{k_r-\omega}{k_r+\omega}\ ^{out}\!f_u^r+\frac{2\sqrt{k_lk_r}}{k_r+\omega}\left[\frac{k_l+\omega}{2k_l}\ ^{out}\!f_v^l+\frac{k_l-\omega}{2k_l} \ ^{out}\!f_u^{l*}\right]\ , \\ \relax
^{in}\!f_H^r=\frac{2\sqrt{k_r\omega}}{k_r+\omega}\  ^{out}\!f_u^r+\left(\sqrt{\frac{k_l}{\omega}}\frac{\omega-k_r}{k_r+\omega}\right)\left[\frac{k_l+\omega}{2k_l}\ ^{out}\!f_v^l+\frac{k_l-\omega}{2k_l} ^{out}\!f_u^{l*}\right]\ ,  \\ \relax
^{in}\!f_H^l=\sqrt{\frac{k_l}{\omega}}\left[\frac{k_l-\omega}{2k_l}\ ^{out}\!f_v^{l*}+\frac{k_l+\omega}{2k_l}\  ^{out}\!f_u^{l}\right].
\eea
Note that there is no contribution to $^{in}\!f_H^l=0$ from the $^{out}\!f_u^r$ modes.
From these the Bogoliubov transformations between the ``in'' and ``out'' creation and annihilation operators can be found as in the previous section. The following expression is found for the number of outgoing created particles in the Unruh state,
\bea
n_u^r&=&\frac{4k_r\omega}{(\omega+k_r)^2}\frac{\delta (0)}{e^{\frac{2\pi\omega}{\kappa}}-1}\Theta(\omega-m_r)\ , \\
n_v^l&=&\frac{1}{4k_l\omega}\left|(k_l+\omega)\frac{\omega-k_r}{\omega+k_r}+(k_l-\omega)e^{\frac{\pi\omega}{\kappa}}\right|^2\frac{\delta (0)}{e^{\frac{2\pi\omega}{\kappa}}-1}\ , \\
n_u^l&=&\frac{1}{4k_l\omega}\left|e^{\frac{\pi\omega}{\kappa}}(k_l+\omega)\frac{\omega-k_r}{\omega+k_r}+(k_l-\omega)\right|^2\frac{\delta (0)}{e^{\frac{2\pi\omega}{\kappa}}-1}+\frac{k_r}{k_l}\left(\frac{k_l-\omega}{k_l+\omega}\right)^2\Theta(\omega-m_r)\ . \ \ \ \ \
\eea
One can verify (again, if wave packets are used then one can set $\delta(0)=1$) that $n_u^r+n_v^l=n_u^l$ above the threshold $\omega>m_r$ while for $0<\omega<m_r$ we have $n_v^l=n_u^l$.

\begin{figure}
\includegraphics[width=6.cm]{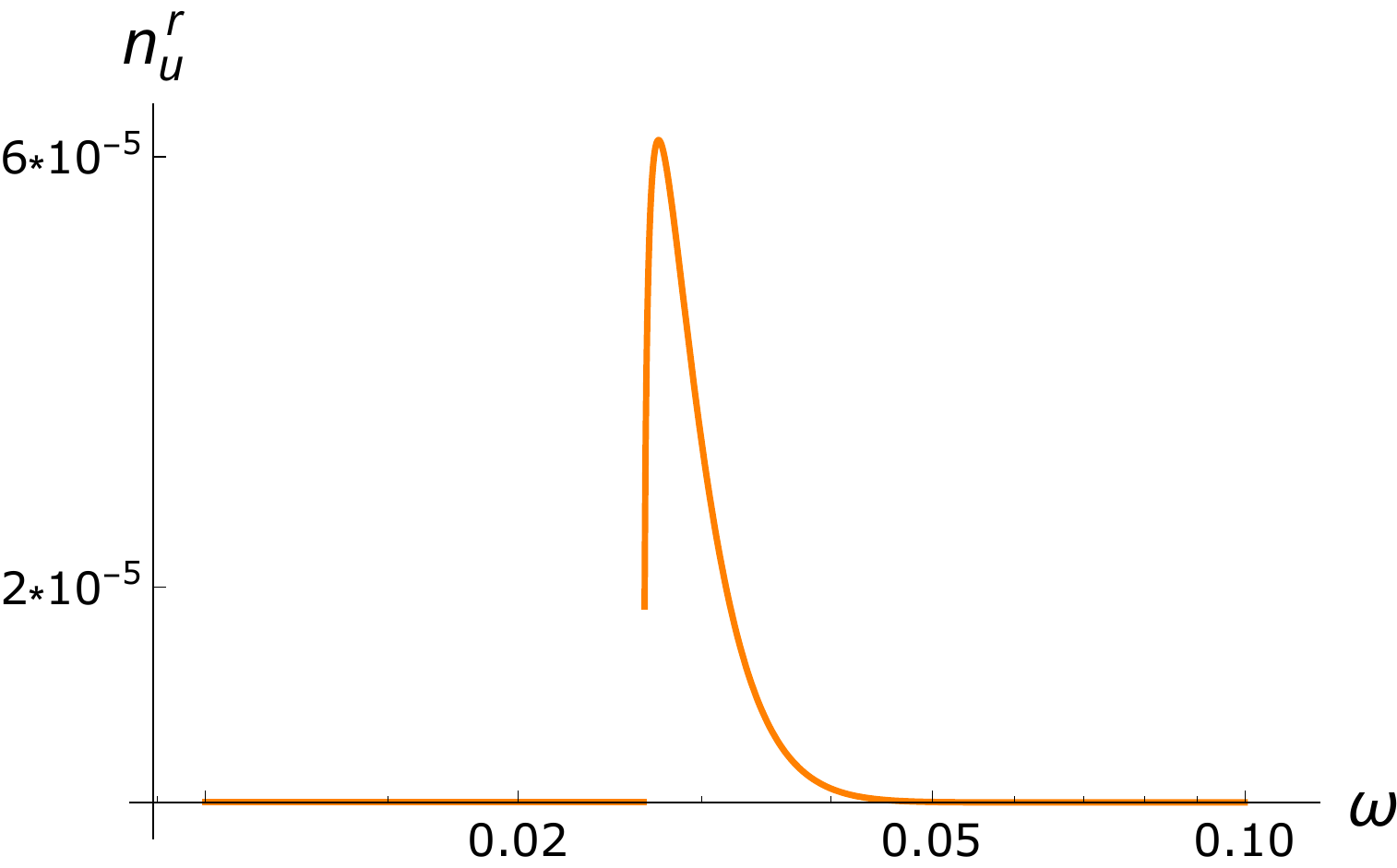}
\includegraphics[width=6.cm]{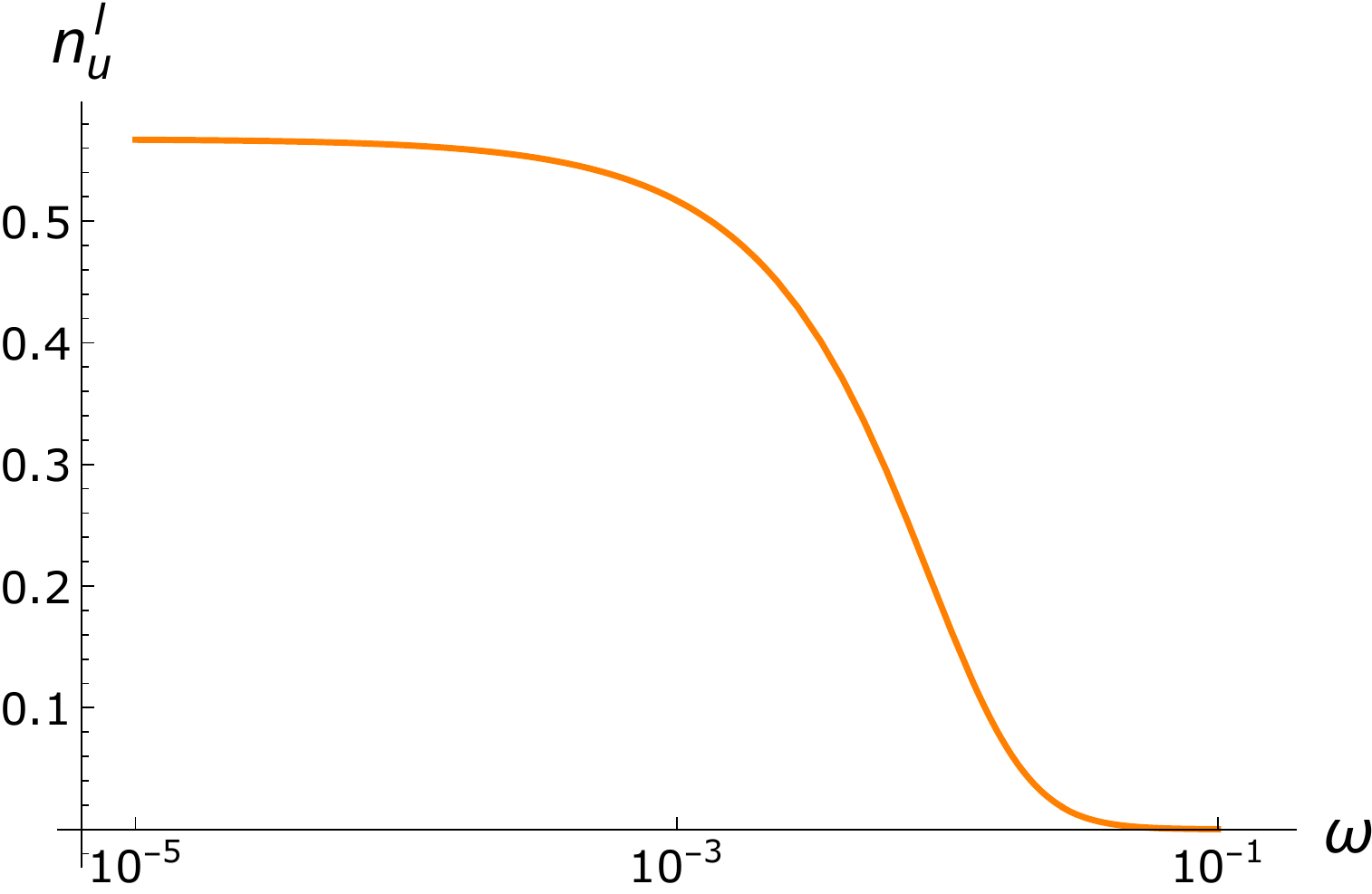}
\caption{\label{Fig:MassComparison} Plots for ${m =4\times 10^{-2}}$.  Left: Plot of $n_u^r$ vs $\omega$.  Note that for ${\omega<m_r}$, $n_u^r = 0$ because no modes reach future null infinity in the $r$ region.  Thus there is a discontinuity in $n_u^r$ at $\w = m_r$.  Right: Plot of $n_u^l$ vs $\omega$ which shows the non-thermal nature of the interior particle number as the low $\omega$ behavior is shown to approach a constant and for large $\omega$ it decays as a power law.  The qualitative behavior of  $n_v^l$ is very similar to $n_u^l$, thus it is not shown. }
\end{figure}

In this toy model we also find that, unlike the exterior region, the emission inside is not thermal. 
Furthermore, $n_u^l$ and $n_v^l$ are finite in the infrared $\omega\to 0$ limit (see also  Ref. \cite{mbh}). 
In the asymptotic interior region,  $  x\to -\infty$, the dispersion relation for the massive modes is $\omega^2-k^2=-m_l^2$ so there is no threshold for the conserved energy, one can have phonons whose energy is below $m_l$, even a zero frequency mode with $|k|=m_l$ exists.   There is a threshold in momentum $|k|>m_l$ for the outgoing $  x\to -\infty$ particles. These features are a consequence of the switching roles between $t$ and $x^*$ inside the BH as we have discussed previously. Note, however, that unlike the energy, momentum is not conserved along the trajectory of the created particle. Finally, the energy of ($u, l$) particles is negative. All of these unusual features exist only inside the BH. The deviation from a thermal spectrum is easily seen in Fig. \ref{Fig:MassComparison}. Note that the spectrum in the exterior is truncated for modes where $\omega<m_R$.

For completeness we can also work out the numbers of created particles in the Boulware state $\left|B\right>$
\bea
N_u^r &=& 0\\\relax
N_u^l&=&N_v^l=\frac{(k_l-\omega)^2}{4\omega k_l}\ .
\eea
One can see that, just like in the previous case, $\left|B\right>$ is no longer a vacuum state in the interior of the BH. Moreover, unlike in the Unruh state the number of created particles diverges as $\omega\to 0$.

\section{Conclusions}

We have investigated scattering in the exterior of the acoustic horizon of a BEC analog BH and anomalous scattering or particle production in its interior in a simple model with massless phonons and a different one for massive phonons. We have considered both the Unruh and Boulware states.  The latter is the natural vacuum state for a static star while the former gives a good approximation in the exterior region to the late time radiation produced by the black hole.  As expected we find for the region outside the horizon that the spectrum at infinity is thermal modulo the graybody factor for the Unruh state and there are no particles for the Boulware state.
In the massive case we find that, as expected, the emitted thermal radiation in the exterior is gapped.

In the interior anomalous scattering produces additional particle production for both massless and massive phonons and this destroys the thermal nature of the spectrum for the Unruh state.  At small frequency the emission is dominated by soft phonons but only in the massless case.
At high frequency one finds that, for the considered models, the particle number falls off like an inverse power of the frequency rather than exponentially.  Not surprisingly particle production also occurs for the Boulware state in the interior.  So the Boulware state remains a vacuum state in the exterior and can only be considered to be an initial vacuum state in the interior.

For massless phonons an unexpected peak was found in the quantities $\w n^\ell_u$ and  $\w n^\ell_v$ when they are plotted as functions of $\w$, with $n^\ell_u$ and $n^\ell_v$ the number of right moving and left moving particles found at future null infinity in the interior.  This peak represents a clear deviation from a thermal spectrum.  It occurs for a limited range of the factors $V_r$ and $V_\ell$ in the delta function potentials~\eqref{2-delta}.

The presence of particle creation 
even for the Boulware state inside a BH can be understood by the fact that the Killing vector $\frac{\partial}{\partial t}$, of which the Boulware modes ``in'' are eigenfunctions, is spacelike inside the horizon. The symmetry associated with it is homogeneity rather  than stationarity. This is clearly seen by the switch of roles of the coordinates $t$ and $x^*$ inside the BH; $x^*$ is timelike  and $t$ is spacelike so a potential depending on $x^*$ is a time dependent potential which as such causes particle creation.

Particle production associated with anomalous scattering induced by curvature and consequent deviation from thermality of Hawking radiation was first noticed by Corley and Jacobson~\cite{coja} in a different context in the region exterior to the event horizon. Specifically, they introduced an ad hoc modification of the two dimensional wave equation for the modes propagating in a BH metric which results in a nonlinear
dispersion relation, subluminal in their case, i.e. $\omega -vk=\pm \sqrt{k^2-\frac{k^4}{k_0^2}}$.   Then they analyzed the influence of the induced anomalous scattering on the spectrum of the particles radiated by the BH in the region exterior to the horizon.  The fact that the anomalous scattering occurs outside the horizon is a peculiar effect of the dispersion they chose. In a genuine General Relativity framework, like the one we use, anomalous scattering and related particle production can occur only inside the horizon, outside the scattering is always the standard one giving just a gray body factor and no extra particle production.

Deviation of thermality of Hawking radiation in the context of BEC analog BHs, where the modification of the relativistic dispersion relation is superluminal, i.e. $\omega -vk=\pm \sqrt{k^2+\frac{k^4}{k_0^2}}$, was first analyzed numerically by Macher and Parentani~\cite{Macher:2009nz}.

Our results are in the context of quantum field theory in curved space,
as such they involve a strictly linear dispersion relation for which there are no superluminal/subluminal modes. The connection to actual analog BHs is that our results should be valid for long wavelength phonons for which the mode equation is approximately the same as that for a massless minimally coupled scalar field in the analog spacetime~\cite{abfp}.
The connection of our results to real black holes is that, in the interior (where the Killing vector is spacelike) the spacetime is dynamic and there is also an effective potential, this time due to the spacetime curvature, and so nonthermal particle production should also occur.

The advantage of analog gravity is that, unlike what happens in the gravitational context, one
 has direct experimental access to the region inside  the horizon and so the spectrum of the phonons emitted there will be  observable.
Our results predict that this spectrum will be completely different from the one emitted outside the horizon.  In particular, it will not be thermal.

\acknowledgments
R. B. would like to thank T. Jacobson, R. Parentani, J. Steinhauer  and S. Weinfurtner for discussion.
A.F. acknowledges partial financial
support by the Spanish Ministerio de Econom'a, Industria y Competitividad Grant
No. FIS2017-84440-C2-1-P, the Generalitat Valenciana Project No. SEJI/2017/042 and the Severo Ochoa Excellence
Center Project No. SEV-2014-0398. R. A. D. thanks
the University of Valencia, where some of this work was done, for hospitality and acknowledges partial financial support from the Paul K. and Elizabeth Cook Richter Memorial Fund.
This work was supported in part by the National
Science Foundation under Grants No. PHY-1505875 and PHY-1912584 to Wake Forest University.

\end{document}